\documentclass[5p,times,preprint]{elsarticle}

\usepackage{lineno}
\usepackage[draft]{hyperref}
\usepackage{listings}
\usepackage{color}
\usepackage{balance}   
\usepackage{graphicx}   
\usepackage{amsmath}     
\usepackage{dingbat}
\usepackage{amssymb}
\usepackage[export]{adjustbox}
\usepackage[autolanguage]{numprint}
\usepackage{tabularx}
\usepackage{mdframed}
\usepackage[]{xcolor}
\usepackage{amsmath}
\usepackage{pifont}
\usepackage{graphicx}
\usepackage{paralist}
\usepackage{float}
\usepackage{makecell}
\usepackage{multirow}
\usepackage{amsfonts}
\usepackage{booktabs}
\usepackage{siunitx}
\usepackage{etoolbox}
\usepackage{subcaption}
\usepackage{xcolor}
\usepackage{listings,setspace,textcomp}

\usepackage{xspace}
\newcommand{\quixbugs}{QuixBugs\xspace}
\newcommand{\junit}{JUnit }

\newcommand{\infiniteloop}{timeout/infinite loop\xspace}
\newcommand{\assertionnotequal}{incorrect output}
\newcommand{\stackoverflow}{stack overflow}
\newcommand{\nullpointer}{null pointer}
\newcommand{\IndexOutOfBounds}{array index error} 
\newcommand{\revision}[1]{\textcolor{black}{#1}}

\newcommand{\bugincorrectlogical}{%A: 
Incorrect logical operator\xspace}
\newcommand{\bugincorrectvariable}{%B: 
Reference to an incorrect variable\xspace}
\newcommand{\bugincorrectcomparison}{%C: 
Incorrect comparison operator\xspace}
\newcommand{\bugmissingbooleanexpression}{%D: 
Missing boolean expression\xspace}
\newcommand{\bugmissingplusone}{%E: 
Missing~`+ 1'\xspace}
\newcommand{\bugmissingminusone}{%F: 
Missing~`- 1'\xspace}
\newcommand{\bugarray}{%G: 
Incorrect array slice\xspace}
\newcommand{\bugcomplex}{%H: 
Missing logic\xspace}
\newcommand{\bugswap}{%I: 
Expression swap\xspace}
\newcommand{\bugincorrectmethodcall}{%J: 
Incorrect method called\xspace} 
\newcommand{\bugwrongconstructor}{%K: 
Wrong constructor call\xspace}
\newcommand{\bugmissingarithmetic}{%L: 
Missing arithmetic expression\xspace}
\newcommand{\bugmissingfunctioncall}{%M: 
Missing function call\xspace}
\newcommand{\bugline}{%N: 
Missing lines with a function call\xspace}
\newcommand{\bugothercodereplacement}{%O: 
Other code replacement\xspace}
\newcommand{\bugmissingassignment}{%P: 
Missing Assignment\xspace}
\newcommand{\bugincorrectarithmeticexpr}{%Q: 
Incorrect arithmetic expression\xspace}
\definecolor{bluekeywords}{rgb}{0.13,0.13,1}
\definecolor{greencomments}{rgb}{0,0.5,0}
\definecolor{redstrings}{rgb}{0.9,0,0}
\usepackage[justification=centering, font=small]{caption}

\usepackage{newfloat}
\usepackage{listings,setspace,textcomp}
\DeclareFloatingEnvironment[fileext=frm,placement={!ht},name=Listing]{listing}

\begin{document}

\begin{frontmatter}

\title{A Comprehensive Study of Automatic Program Repair on the QuixBugs Benchmark}

%% Group authors per affiliation:

\author[kth]{He Ye\corref{mycorrespondingauthor}}
\cortext[mycorrespondingauthor]{Corresponding authors}
\ead{heye@kth.se}

\author[france]{Matias Martinez}
\ead{matias.martinez@uphf.fr }

\author[kth]{Thomas Durieux}
\ead{thomas@durieux.me}

\author[kth]{Martin Monperrus}
\ead{martin.monperrus@csc.kth.se}

\address[kth]{KTH Royal Institute of Technology, Sweden}
\address[france]{Universit\'e  Polytechnique Hauts-de-France, France}

\begin{abstract}
Automatic program repair papers tend to repeatedly use the same benchmarks. This poses a threat to the external validity of the findings of the program repair research community.
In this paper, we perform an empirical study of automatic repair on a benchmark of bugs called QuixBugs, which has been little studied. In this paper, 1) We report on the characteristics of QuixBugs; 2) We study the effectiveness of 10 program repair tools on it; 3) We apply three patch correctness assessment techniques to comprehensively study the presence of overfitting patches in QuixBugs. Our key results are:
\begin{inparaenum}[1)]
\item  16/40  buggy programs in \quixbugs{} can be repaired with at least a test suite adequate patch;
\item  A total of 338 plausible patches are generated on the \quixbugs  by the considered tools, and 53.3\% of them are overfitting patches according to our manual assessment;
\item The three automated patch correctness assessment techniques, $RGT_{Evosuite}$, $RGT_{InputSampling}$ and $GT_{Invariants}$, achieve an accuracy of 98.2\%, 80.8\% and 58.3\% in overfitting detection, respectively.
\end{inparaenum}
To our knowledge, this is the largest empirical study of automatic repair on \quixbugs, combining both quantitative and qualitative insights.
All our empirical results are publicly available on GitHub in order to facilitate future research on automatic program repair.
\end{abstract}


\end{frontmatter}

\section{Introduction}

% context about the research
Automatic program repair aims to provide  fixes to software bugs in an automated way. Test suite based repair, notably introduced by GenProg \cite{genprog}, is a widely studied family of techniques in program repair. In test suite based repair, test suites are used as an executable specification of the program, with at least one failing test that reveals the bug. Test suite based repair can be further divided into generate-and-validate techniques and synthesis-based techniques. Generate-and-validate techniques, such as GenProg \cite{genprog}, Astor \cite{Astor2019Journal}, CapGen \cite{capgen-ICSE18}, first generate as many patches as possible and then use the test suite to validate if the patch makes all tests pass. On the other hand, synthesis-based techniques such as AutoFix \cite{autofix}, SemFix \cite{semfix}, and Nopol \cite{Nopol} first extract constraints based on test suite execution and then synthesize a patch \cite{martinbibli,TSE-repair-survey}.

% problem addressed / motivation
Recent automatic program repair papers tend to repeatedly use the same benchmarks.
In program repair for C code, the ManyBugs \cite{manybugs} benchmark or its derivative is dominant.
In the context of program repair for Java, Defects4J \cite{defects4jbenchmark} is used in almost all evaluations of recent program repair approaches, including recently \cite{sketchfix,capgen-ICSE18,patch-correctness-icse18}.
However, repeatedly using the same benchmarks  poses a threat to the external validity of the community's knowledge. 
% why the problem is important and significant
The main threat is that the improvement that we now observe in the literature may only be valid for the benchmark under consideration but would not hold for other benchmarks.
Even worse, those claimed improvements, if they only hold on the benchmark, maybe decorrelated from for real usages by practitioners. Fortunately, the importance of external validity is acknowledged by many researchers.

\emph{Problem: Research on program repair tends to repeatedly use the same benchmarks. This is a threat to the external validity of the results of our research community.}

As building sound and conclusive empirical knowledge is key to science, reducing this major threat of external validity in the context of program repair is the main motivation of this paper.
To reduce the threat, we aim at doing a empirical program repair study on a new and well-formed bug benchmark.

% what is the solution, what we do in this paper
In this paper, we perform an automatic repair empirical study on a  benchmark called QuixBugs which was recently presented by Lin et al.~\cite{quixbugs-orig}.
QuixBugs is a program repair benchmark with 40 buggy algorithmic programs specified by test cases. 
The buggy programs are both available in Python and Java.
\revision{
In this paper, we conduct the following four experiments on Quixbugs:
}
\begin{inparaenum}[1)]
\item We prepare QuixBugs for automatic program repair in Java;  
\item We select ten representative test suite based repair tools,  Arja~\cite{Yuan2017ARJAAR}, Cardumen~\cite{cardumen}, Dynamoth~\cite{dynamoth},
JGenProg~\cite{Astor2019Journal}, JMutRepair~\cite{Astor2019Journal}, JKali~\cite{Astor2019Journal}, 
Nopol~\cite{Nopol}, NPEFix~\cite{NPEFixAR},  Tibra~\cite{Astor2019Journal}, and the Java implementation of RSRepair \cite{Qiissta15}, and execute them over all  buggy programs of QuixBugs. This results in 16/40 buggy programs being repaired by 338 different plausible patches; 
\revision{
\item We perform manual assessment for the generated plausible patches and  manually classify them as 158 correct patches and 180 overfitting patches;}
\item We assess the correctness of the plausible patches by three automated patch correctness assessment techniques: $RGT_{Evosuite}$, $RGT_{InputSampling}$ and $GT_{Invariants}$. 
\revision{
We compute the accuracy of these three automated techniques are 98.2\%, 80.8\% and 58.3\%, respectively. }
\end{inparaenum} 
% what is the value of our solution, of our experiment
This novel empirical study on a benchmark never used in a program repair context provides valuable findings that improve the external validity of program repair research. 
Our empirical study sets a baseline for future research of automatic program repair on QuixBugs. 

To sum up, our contributions are:
\begin{itemize}

\item A new version of \quixbugs{} that is usable for automatic repair research on Java programs, together with extensive data about the characteristics of \quixbugs{}.

\item The confirmation of two empirical facts of program repair, improving their external validity:
\begin{inparaenum}[\it 1)]
\revision{
\item Our manual assessment shows that 53.3\% of generated patches are overfitting, this confirms that the state-of-the-art of program repair tools produces a large number of overfitting patches \cite{overfit-problem, cure-worse-15,LeICSE18};
}
\item Our empirical study shows the considered automatic program repair tools are able to correctly repair seven buggy programs, this confirms the state-of-the-art program repair tools also produce correct patches \cite{Qiissta15,Defects4JExperiment}.
\end{inparaenum}

\item Three new and important findings about automatic program repair:
\begin{inparaenum}[\it 1)]
\item Certain program repair tools are able to repair programs with only failing test cases and no passing tests at all;
\item \revision{It is feasible and effective to use automated patch assessment techniques to identify overfitting patches with an accuracy of up to 98.2\%;} 
\item Invariants based patch assessment suffers from a large number of false positives. 
\end{inparaenum}

\item Experimental data that is made publicly available for facilitating future research \cite{ourrepo}. Our 338 plausible patches on \quixbugs{} and their correctness labels are consolidated for future studies on program repair.

\end{itemize}

%%new text: explanation of the diff with the workshop %%
This paper supersedes a previous version \cite{quixbugs} presented at the \textit{International Workshop on Intelligent Bug Fixing}.  In comparison, this article makes the following extensions.
The program repair empirical study involves ten repair tools (expanding from five in the previous version).
This new work presents and discusses the 338 plausible patches versus only 64 patches discussed in the previous version.
This study considers a third automated patch assessment technique based on invariants. To our knowledge, this technique has only been studied by Yang and Yang \cite{ibf20-invariant}, and at a smaller scale (our dataset of patches is three times larger than that of \cite{ibf20-invariant} -- 338 versus 96).
This journal extension provides novel results that compare the accuracy of three assessment techniques and discuss the false positive problem of invariants based patch correctness assessment, both of which have never been reported before.

The remainder of this paper is organized as follows. \autoref{benchmark} presents how we prepare a new version of QuixBugs for the usage of automatic repair for Java programs. \autoref{experiment} presents four research questions (RQs) of our study and corresponding methodologies for these  RQs. \autoref{results} 
presents our empirical results to answer the RQs. \autoref{threat} analyzes the threat of our study. 
\revision{\autoref{discussisons} discusses the new findings of using QuixBugs and the future improvements for program repair tools. }
\autoref{relatedwork} compares the related work of our study and we conclude our study in \autoref{conclusion}.

\section{Benchmark preparation}
\label{benchmark}

%1) Description (i.e., the mentioned points: subjects, json as oracle, engine that run programs, etc) 
\quixbugs by Lin et al.~\cite{quixbugs-orig} is a benchmark of 40 bugs from 40 classic algorithms such as sorting algorithms of \textit{bucket sort, merge sort and quick sort}.
All bugs of \quixbugs were collected from the \textit{Quixey Challenges} \cite{quixey-challenge}, which consisted of giving human developers one minute to fix one program with a bug on a single line. 
The original \quixbugs benchmark contains: 
\begin{inparaenum} [\it 1)]
\item A set of 40 buggy programs available both in Python and in Java; 
\item For 31 out of 40 programs: JSON files with a set of inputs and expected outputs for each program;
\item An engine that takes a program name, executes the program using the inputs from the corresponding JSON file, and prints the expected and obtained output; 
\item For the remaining 9 out of 40 programs, a Java class that has encoded the inputs and outputs and prints the obtained output.
\end{inparaenum}

However, the initial version of \quixbugs was not usable for doing  automatic program repair in Java. 
Monperrus \cite{martinbibli} states that, in the context of test suite based repair, a ``usable" benchmark must have \cite{martinICSE14}:
\begin{inparaenum}[\it 1)]
\item A clear, explicit, and not biased construction methodology;
\item Regression oracles. For test suite based repair approaches such as GenProg \cite{genprog} the oracles are the test suites:  
the failings test cases are the bugs oracles and assert the presence of a bug, while
the passing test cases are the regression test cases that assert the correctness of the program w.r.t the inputs-outputs encoded in the test suite;
\item Real bugs (i.e., not seeded).
\end{inparaenum}

%2) Problematic 
Unfortunately, the initial version of \quixbugs does not fulfill some of the aforementioned criteria.
We summarize the problems of the initial version of \quixbugs as: 
\begin{inparaenum} [\it 1)]
\item It did not provide any regression oracle, \revision{this not fulfill the second requirement of a usable benchmark  (bugs and regression oracles);}
\item Programs contained compilation errors (for 5 programs), \revision{this does not satisfy the first requirement of a usable benchmark (a clear,  explicit  and not biased construction methodology);}
\item Incorrect values to test buggy programs (for 3 programs), \revision{which also not fulfill the first requirement of a usable benchmark;}
\item Missing test assertions (for 9 programs), \revision{this violates the second  requirement of a usable benchmark;}   
\item Missing a ground truth Java  version (for all programs), \revision{without the ground truth patches provided, the correctness assessment of generated patches is harder. }
\end{inparaenum}

To overcome the mentioned limitations that hamper its use by test suite based repair approaches, we introduce a new version of \quixbugs supplemented with test cases for reproducing buggy behaviors and a  ground truth version for evaluating automatic repair patches. 
This new version of \quixbugs was already peer-reviewed and accepted by the \quixbugs authors and integrated to their public repository at GitHub.
The steps we carried out for creating the new version are:
\begin{inparaenum}[\it 1) ]

% // pb1 compilation error
\item Fix uncompilable Java programs. By compiling the initial version Java programs of QuixBugs, we noticed that there were compile errors in some programs (e.g.,
%%%Hack to break the line:
\emph{BREADTH\_}\emph{FIRST\_}   \emph{SEARCH}). 
Some compilation errors were designed as part of buggy programs. 
However, most automatic repair tools do dynamic analysis of buggy programs. Hence, we need them all to be compilable and able to run the original buggy programs. 

% // pb2 incorrect values
\item Fix incorrect test data. To test 31 out of 40 buggy Java programs, \quixbugs provides pairs of inputs and expected outputs written in JSON files.
However, we found that some expected outputs from programs \textit{KNAPSACK}, \textit{SQRT} and \textit{PASCAL} were incorrect.
Once we detected all incorrect inputs and outputs, we corrected them. 

% pb3 JSON solution2: Junit generation
\item Creation of JUnit tests from JSON files. \quixbugs uses a specific test driver based on JSON test cases. It executes the program using the inputs, and prints both the expected and actual outputs. 
However, automatic repair tools usually expect \junit tests as oracle specifications: each test executes the program passing the inputs via parameters and then compares the obtained  output  with that one expected via assertions.
% solution
Thus, we implement an automatic \junit test generator to generate JUnit tests from the JSON files. 
In total, we generated 224 JUnit tests (test methods in  JUnit) for the 31 programs having their inputs-outputs encoded in the JSON files.

% // manual rewrite of the unconventional main-based tests
\item Creation of JUnit tests from ad-hoc assertion-less tests. There are 9 out of 40 Java programs from \quixbugs that are tested through a simple ad-hoc main method that starts with encoded inputs, calls the program using them as arguments, and finally prints the obtained output. 
This method is not usable by a test suite based program repair tool. 
Thus, we have manually rewritten those methods to produce 35 JUnit tests for these 9 programs.
In total, our preparation has resulted in 259 \junit test methods over 40 programs.

% // manual correct java programs
\item Creation of ground truth Java  programs. By default, the \quixbugs does not provide a ground truth version for the Java buggy programs.
Automatic program repair researchers need those ground truths to compare them with the generated patches to assert their correctness.
We created  ground truth versions based on those provided by  \quixbugs originally written in Python. 
%the provided correct Python version from \quixbugs.
\end{inparaenum}

To summarize, \quixbugs was initially not usable for automatic repair tools in Java. In this section, we presented the tasks we carried out to build a new version of \quixbugs that can be used to evaluate the effectiveness of the test suite adequate repair tools. The new version of \quixbugs contains \junit test oracles and ground truth programs, it was public peer-reviewed by the \quixbugs authors, organized with Travis and Gradle components. 
All those changes have already been contributed to the research community on the \quixbugs repository.%: https://github.com/jkoppel/QuixBugs

%%%%%%%%%%%%%%%%%%%

%%%%%%%%Experiment start%%%%%%%%%%%
\section{Empirical Study}
\label{experiment}
We now present our empirical study on the effectiveness of test suite based repair approaches on the \quixbugs benchmark. The empirical study covers  several dimensions of automatic repair: benchmark analysis, repair effectiveness, patch correctness assessment. First, we list the research questions (RQs) of our work, we then describe the research methodology for each RQ. 
% Finally, we present our experiment environment details.

\subsection{Research Questions}
For this empirical study on program repair for QuixBugs, we pose the following research questions (RQs):
\begin{itemize}   
\item RQ1: What are the main characteristics of the \quixbugs benchmark?

\item RQ2: How many buggy programs of \quixbugs can be automatically repaired with test suite adequate patches?
\revision{\item RQ3: To what extent are the generated patches for QuixBugs correct?}
\revision{\item RQ4: To what extent do automated patch assessment techniques accurately classify overfitting patches?}
\end{itemize}   

In RQ1, we are interested in the statistics of \quixbugs, including the type of bug, lines of code (LOC), \junit tests, code coverage, etc. In RQ2, we consider one kind of automatic repair called test suite based repair. 
In test suite based repair, a bug is said to be repaired if a patch makes all tests pass. In that case, such a patch is called \emph{test suite adequate patch} or \emph{plausible patch}.
We focus on how many test suite adequate patches could be generated by the state-of-the-art test suite based repair approaches. 
\revision{In RQ3, we conduct a manual assessment to evaluate how many patches generated in the experiment of RQ2 are actually correct. Finally, in RQ4, we study the effectiveness of three techniques to automatically classify correct and overfitting patches, and we compare their results with those from the manual assessment.}

\subsection{Protocols}
This section presents the protocols of our empirical study of automatic program repair on QuixBugs.

\newcommand\RQexploration{RQ1: \quixbugs Benchmark Analysis}
\subsubsection{\RQexploration}
Bug understanding \cite{survey-bugs} is important for designing program repair tools and to analyzing the effectiveness of those tools.
For each buggy program of \quixbugs, we gather and compute the following information:

\paragraph{Types of bugs}
The previous research reports the existence of the observational correlation between the bug fix and the cause of bugs \cite{relation-bug-cause}. 
In our study, we collect and present the type of bug. \quixbugs contains various types of bugs such as incorrect comparison operators, incorrect array slice, etc. 
This allows us to analyze the capability of the program repair tools to repair buggy programs and to determine the most repaired bug types.

\paragraph{Numerical characteristics}
We compute numerical characteristics: the lines of code (LOC) of the program, the number of passing \junit tests, failing \junit tests, the test execution time and branch coverage. 
We rely on Cobertura\footnote[1]{Cobertura website: \url{http://cobertura.github.io/cobertura/} (visited \today)} to calculate the branch coverage for each program.

\paragraph{Input domain}
We extract the program preconditions and the input domain of each program. The program preconditions are constraints for the input domain. We discuss this to remind the future work on QuixBugs to sample tests that should be aware of the program preconditions. 

\paragraph{Failures types}
We manually collect the failure symptoms when executing test cases for each buggy program of \quixbugs dataset. A bug can produce:
\begin{inparaenum}[\it 1)]
\item An incorrect output that triggers an assertion fail;
\item An error in the execution (e.g., array index error);
\item An exception thrown by the program (e.g., null pointer exception or stack overflow);
\item A \infiniteloop. 
\end{inparaenum}

\paragraph{Unique characteristics}
We discuss the unique characteristics of the \quixbugs dataset compared with the benchmarks from the literature. 

\newcommand\RQrepairability{RQ2:  Repairability of QuixBugs}
\subsubsection{\RQrepairability} 

To conduct our program repair empirical study on \quixbugs, we first select appropriate program repair tools.
For this, we consider three criteria: 
\begin{inparaenum}[\it 1)]
\item The repair tool must handle Java programs as \quixbugs{} programs are written in this programming language\footnote{Note, \quixbugs also contains Python implementation of those bugs, however we focus on the Java implementations since few repair tools are available for Python.}; 
\item The repair tool must implement a test suite based repair approach; 
\item  The repair tool must be publicly-available and continuously maintained.
\end{inparaenum} 

According to these criteria, we finally select ten of program repair tools: Arja~\cite{Yuan2017ARJAAR}, JGenProg~\cite{Astor2019Journal}, JKali~\cite{Astor2019Journal}, JMutRepair~\cite{Astor2019Journal}, Cardumen~\cite{cardumen}, Tibra~\cite{Astor2019Journal}, Nopol~\cite{Nopol}, Dynamoth~\cite{dynamoth}, NPEFix~\cite{NPEFixAR} and the Java implementation of  RSRepair~\cite{Qiissta15} by~\cite{Yuan2017ARJAAR}.
The ten repair tools target Java programs, are test suite based, and are publicly available on GitHub. All the ten repair tools take as input the source code of a buggy program and the corresponding test suite which contains at least one failing test case, and generate, when it's possible, one or more patches that make all test cases pass.
We combine the patches generated during our empirical study in \cite{quixbugs-orig} with patches generated from our recent work \cite{RepairThemAll2019}.
\revision{Each of the ten repair tools has been executed on all QuixBugs programs. }
We do not stop the repair process after finding the first patch, and we consider all generated patches, even if there are several patches for the same bug. 

We carefully record and discuss:
\begin{inparaenum}[\it 1)]
\item The number of bugs that are repaired by the considered 10 systems; 
\item The bug types of the repaired programs;
\item How test cases impact the repair tools; 
\item The test failure symptoms of repaired programs.
\end{inparaenum}

%% methodology RQ3
\newcommand\RQmanualassessment{RQ3: Manual Patch Correctness Assessment}

\subsubsection{\RQmanualassessment}
\revision{
Previous works have shown that program repair tools tend to generate a large number of overfitting patches (i.e., flawed repairs). 
In our work, per the previous terminology \cite{cure-worse-15,patch-correctness-icse18,issta17-difftgen}, we use the term overfitting to refer to those patches that pass all human-written test cases (i.e., test suite adequate patch) but still do not correctly repair the bug.
Those flawed repairs are produced because of the weaknesses of the test suite used as an oracle, which is not able to completely specify the expected program behavior.
To assess the correctness of patches generated for Quixbugs' buggy programs, we perform the manual assessment as previous researchers have done on other benchmarks \cite{ACS,capgen-ICSE18,Defects4JExperiment,Yuan2017ARJAAR}. 
We manually compare the automatically generated patches with the human-written patches. 
If a generated patch is identical or semantically equivalent (i.e., the actual behavior is the same) to the human-written patch, it is considered as correct.
Otherwise, a patch is deemed as overfitting if
\begin{inparaenum}[\it 1)]
\item it does not/partially fix the bug, or
\item it introduces a new bug.
\end{inparaenum}
To overcome the bias of manual assessment, all results are discussed among at least two authors. 
Our evaluation of patch correctness is publicly available on our GitHub repository \cite{ourrepo}.
}

%% methodology RQ4
\newcommand\RQcorrectnessassessment{RQ4: Automated Patch Correctness Assessments}
\subsubsection{\RQcorrectnessassessment}
\label{q3methodology}

\revision{As shown in previous research \cite{patch-correctness-icse18,drr}, manual assessment of program repair patches is a hard, time consuming and biased task.} 
Thus, we also consider three automated patches correctness assessment techniques to identify overfitting patches, proposed by previous research:
\begin{inparaenum} [a)]
\item Using automatically generated tests based on a ground truth version (i.e., the human-written patch) \cite{zhongxingpaper};
\item Using automatically generated tests by a program specific generator based on a ground truth version \cite{randomtest};  
\item Using dynamic program invariants based on a ground truth version \cite{ibf20-invariant}.
\end{inparaenum}
We now describe how each of those techniques works.

% using automated testing
\paragraph{Search-based test generation for patch assessment}
Using automated test generation is one way for assessing patch correctness \cite{patch-correctness-icse18,issta17,zhongxingpaper,Opad}. 
The idea of this technique is to generate new test cases that complement the already provided (potentially incomplete) test suite.
In this paper, we consider Evosuite \cite{evosuite}, a state-of-the-art automated test generation tool, for generating those new correctness assessment tests. We have chosen Evosuite according to the results of \cite{Shamshiriase15,drr}, which have shown that Evosuite is the most effective tool for this usage. 
The search-based test generator technique takes as input a ground truth program that is used as oracles, which means that the outputs from the ground truth programs on  given inputs are the expected outputs (i.e., oracles), including both values and exceptions.
Per the previous terminology, this patch assessment technique is named $RGT_{Evosuite}$, which is based on the ground truth programs for test generation.

For each of the 40 buggy programs in the \quixbugs{} dataset, we invoke Evosuite a fixed number $n$ of times. Eventually, we obtain $n$ different independent \junit test suites for each program.
Since Evosuite is a randomized algorithm, we take $n=30$ per the recommended practice \cite{randomness-guide,zhongxingpaper}.
We always remove those generated tests that fail on the ground truth version, because they are ill-formed for our task. 
We execute these generated tests over the patches generated by the repair tools.
\revision{In the assessment of $RGT_{Evosuite}$, a patch is assessed as \textit{overfitting} if it makes any automatically generated  Evosuite test cases fail.}
If no generated test fails, we consider that the correctness of the  patch is \textit{unknown} (and not \textit{correct} because the generated tests only sample the input domain, they do not assess the behavior over the full input domain).

\paragraph{Program specific test generation}
We consider a second random testing approach called $RGT_{InputSampling}$, which randomly samples the test inputs based on the ground truth programs.
$RGT_{InputSampling}$ is an implementation of random testing \cite{randomtest} for \quixbugs.
It samples the input space according to a specification of the input space, a uniform distribution for sampling and it uses the ground truth version as oracles \cite{mechtaev2018semantic}. 
If the ground truth version throws an exception on a generated input, the input is considered as invalid, the input is discarded. 
For implementing $RGT_{InputSampling}$, we manually identify the domain of each input variable for each program in \quixbugs.
The test generator is configured to sample the input space with the goal of getting a fixed number of valid test cases with no exception. 
For each program in \quixbugs, we generate 300 test cases with $RGT_{InputSampling}$. 
We run those test cases on all generated patches of \quixbugs programs.
\revision{
In the assessment of $RGT_{InputSampling}$, a patch is assessed as \textit{overfitting} if it makes any randomly generated program specific test cases fail.}

\paragraph{Invariants detection for patch assessment}
We consider a third automated patch assessment based on invariants captured from ground truth program execution, such assessment technique is called ground truth invariants, aka, $GT_{Invariants}$.
An invariant is a property that holds at a certain point or points in a program.  
A program point is a specific place in the source code, such as immediately before a particular line of code.  
Invariants detection runs a program, observes the values that the program computes, and then reports properties that were true over the observed program executions.
The invariants based patch assessment technique first infers program invariants from the ground truth version and uses them to assess overfitting patches, per the technique of \cite{ibf20-invariant}.
It executes the patched programs based on the provided manual tests and checks whether all inferred invariants still hold.
\revision{
In the assessment of $GT_{Invariants}$, a patch is assessed as \textit{overfitting} if it violates any invariants hold for ground truth program executions.
To capture invariants in the ground truth programs and checking whether they hold for the captured programs, we use the tool Daikon \cite{daikon}}.

\newcommand\RQfalsepositive{Accuracy of automated patch assessment}

\revision{
\paragraph{\RQfalsepositive}
To evaluate the accuracy of automated patch assessments, we compare the automated patch assessment results with manual assessment, where manual assessment is considered as ground truth.}
Specifically, we compute the corresponding false positives and true negatives as follows:
\begin{itemize}
\item{True Positive (TP): a patch classified as overfitting by manual assessment is also classified as overfitting by an automated assessment.}
\item{False Positive (FP): a patch classified as correct by manual assessment is classified as overfitting by an automated assessment.}
\item{True Negative (TN): a patch classified as correct by manual assessment is classified as correct by an automated assessment.}
\item{False Negative (FN): a patch classified as overfitting by manual assessment is classified as correct by an automated assessment.}
\end{itemize}
Finally, the accuracy of an assessment technique is computed with the following evaluation formula:
\begin{equation}
\label{equation:accuracy}
     Accuracy = \dfrac{TP+TN}{(TP+TN+FP+FN)}
\end{equation}

\begin{table*}[!ht]
\caption{Descriptive Statistics about the QuixBugs Benchmark.}
\label{tab:quixbug-tests}
\footnotesize
\centering
\renewcommand{\arraystretch}{1.28}
    
\begin{tabularx}{0.98\textwidth}{@{}Xlrrrrrl@{}}       

\toprule[0.2ex]	
\multirow{2}{*}{Buggy Program Name}& \multirow{2}{*}{Bug Type}&\multirow{2}{*}{LOC} & Passing&Failing&Code&Exe&\multirow{2}{*}{Failure Symptoms} \\

& & & Tests&Tests&Coverage&Sec.&\\
\toprule[0.2ex]
BITCOUNT&\bugincorrectlogical&10&0&9&100\%&900&\infiniteloop \\
\hline 
BREADTH\_FIRST\_SEARCH&\bugmissingbooleanexpression&30&4&1&100\%&$<$1&\IndexOutOfBounds \\
\hline                 
BUCKETSORT&\bugincorrectvariable&17&0&6&100\%&$<$1&\assertionnotequal \\
\hline    
DEPTH\_FIRST\_SEARCH  &\bugline&23&4&1&100\%&$<$1&\stackoverflow\\
\hline    
DETECT\_CYCLE&\bugmissingbooleanexpression&17&4 &1&100\%&$<$1& \nullpointer \\
\hline    
\multirow{2}{*} {FIND\_FIRST\_IN\_SORTED }  & \multirow{2}{*}{\bugincorrectcomparison}& \multirow{2}{*}{22}& \multirow{2}{*}{4} & \multirow{2}{*}{3}&\multirow{2}{*}{90\%}& \multirow{2}{*}{120}&\infiniteloop{} (1) \\
 &&&&&&&\IndexOutOfBounds{} (2) \\
\hline    
FIND\_IN\_SORTED &\bugmissingplusone&19&5&2&100\%&$<$1&\stackoverflow \\
\hline
FLATTEN &\bugmissingfunctioncall&18&1&6&83\%&$<$1&\stackoverflow \\
\hline    
GCD&\bugswap &10&0&5&100\%&$<$1&\stackoverflow \\
\hline    
GET\_FACTORS &\bugwrongconstructor&17&1&10&100\%&$<$1&\assertionnotequal \\
\hline  
HANOI&\bugincorrectvariable&53&0&7&100\%&$<$1&\assertionnotequal\\
\hline          
IS\_VALID\_PARENTHESIZATIONl &\bugothercodereplacement&15&2&1&100\%&$<$1&\assertionnotequal \\
\hline    
KHEAPSORT&\bugmissingfunctioncall&29&1&3&100\%&$<$1&\assertionnotequal \\
\hline    
KNAPSAC&\bugincorrectcomparison&30&4&6&100\%&2&\assertionnotequal\\
\hline    
KTH &\bugincorrectvariable&25&3&4&100\%&$<$1&\IndexOutOfBounds \\
\hline    
%lcs\_length &\bugarray&48&1&8&95\%&$<$1&\assertionnotequal \\ 

\multirow{2}{*}{LCS\_LENGTH}  &\bugarray&\multirow{2}{*}{48}&\multirow{2}{*}{1}&\multirow{2}{*}{8}&\multirow{2}{*}{95\%}&\multirow{2}{*}{$<$1}& \multirow{2}{*}{\assertionnotequal }\\
   &\bugmissingbooleanexpression&&&&&&  \\
 \hline 
LEVENSHTEIN  &\bugmissingplusone&15&1&6&100\%&$<$1&\assertionnotequal\\
 \hline    
 LIS  &\bugcomplex&27&0&4&91\%&$<$1&\assertionnotequal\\
\hline    
LONGEST\_COMMON\_SUBSEQUENCE &\bugmissingfunctioncall&14&6 &4&91\%&$<$1&\assertionnotequal \\
\hline    
MAX\_SUBLIST\_SUM &\bugmissingfunctioncall&13&2&4&100\%&$<$1&\assertionnotequal\\
\hline    
MERGESORT &\bugincorrectarithmeticexpr&40&0 &12&100\%&$<$1&\stackoverflow\\
\hline    
MINIMUM\_SPANNING\_TREE&\bugcomplex&67&0&3&72\%&$<$1&concurrent modification \\
\hline    
NEXT\_PALINDROME&\bugmissingminusone&28&4&1&87\%&$<$1&\assertionnotequal \\
\hline    
NEXT\_PERMUTATION  &\bugincorrectcomparison&32&0&8&83\%&$<$1&\assertionnotequal \\
\hline               
\multirow{2}{*}{PASCAL}  &\multirow{2}{*}{\bugmissingplusone}&\multirow{2}{*}{29}&\multirow{2}{*}{1}&\multirow{2}{*}{4}&\multirow{2}{*}{100\%}&\multirow{2}{*}{$<$1}& \IndexOutOfBounds{} (3)\\
   &&&&&&& \assertionnotequal{} (1) \\
\hline    
POSSIBLE\_CHANGE &\bugmissingbooleanexpression&23&0&9&100\%&$<$1&\IndexOutOfBounds \\
\hline    
POWERSET  &\bugcomplex&24&1&4&100\%&$<$1&\assertionnotequal\\
\hline    
QUICKSORT  &\bugincorrectcomparison&37&12&1&87\%&$<$1&\assertionnotequal \\
\hline    
REVERSE\_LINKED\_LIST &\bugmissingassignment&12&1&2&100\%&$<$1&\nullpointer \\
\hline    
RPN\_EVAL&\bugswap&28&3&3&100\%&$<$1&\assertionnotequal \\
\hline    
SHORTEST\_PATH\_LENGTH &\bugothercodereplacement&49&2&2&92\%&$<$1&\assertionnotequal\\
\hline    
SHORTEST\_PATH\_LENGTHS  &\bugswap&31&0&4&100\%&$<$1&\assertionnotequal\\
\hline    
SHORTEST\_PATHS &\bugmissingfunctioncall&55&0&3&100\%&$<$1&\assertionnotequal \\
\hline    
SHUNTING\_YARD&\bugline&31&0&4&100\%&$<$1&\assertionnotequal\\
\hline    
SIEVE &\bugincorrectmethodcall&35&1&5&75\%&$<$1&\assertionnotequal\\
\hline    
SQRT&\bugincorrectarithmeticexpr&9&1&6&100\%&360&\infiniteloop\\
\hline    
SUBSEQUENCES&\bugline&22&2&12&100\%&$<$1&\assertionnotequal \\
\hline    
TO\_BASE &\bugswap&14&0&7&100\%&$<$1&\assertionnotequal\\
\hline    
TOPOLOGICAL\_ORDERING&\bugincorrectmethodcall&25&0&3&100\%&$<$1&\assertionnotequal \\
\hline    
WRAP &\bugline&22&0&5&75\%&$<$1&\assertionnotequal \\
  \bottomrule[0.2ex]
Total&-& \numprint{1034} &70&189& - &-&-\\
 \bottomrule[0.2ex]
\end{tabularx} 
\renewcommand{\arraystretch}{.9}
% \resizebox{1\textwidth}{!}{
%\begin{tabularx}{0.9\textwidth}{@{}X@{}}  
%\bottomrule[0.2ex]
%\end{tabularx}
\end{table*}

%%%%%%%%%%%%result start %%%%%%%%%%%%%
\section{Empirical Results}
\label{results}
We now present and discuss the empirical results of our four research questions.

\subsection{Results for \RQexploration}
\label{quixbugresult}

% program size
\autoref{tab:quixbug-tests} presents the characteristics of \quixbugs, including the numerical statistics (e.g., LOC) and failure symptoms (e.g., incorrect output, null pointer exception). 
Program names are given in the first column in alphabetical order. 

%A:Incorrect Assignment operator, 
%B:Incorrect variable, 
%C:Incorrect comparison operator,  \\
%D:Missing condition, E:Missing/added+1, F:Variable swap, G:Incorrect array slice, H:Variable prepend   \\
%I:Incorrect data structure constant, J:Incorrect method called, K:Incorrect field dereference, \\
%L:Missing arithmetic expression, M:Missing function call, 
%N:Missing line\\

\revision{
\paragraph{Type of bugs}
The second column presents the type of bug in each program. 
There are 17 different bug types.
The most frequent bug types on \quixbugs{} are:
\begin{inparaenum}[\it 1)]
\item \emph{\bugmissingfunctioncall} in five programs.
In those buggy programs, function invocations are missing. 
This means the patch that repairs this type of bug typically adds a function invocation. 
For example, to repair the bug for \textit{FLATTEN}, the existing variable \textit{x} should be replaced by \textit{flatten(x)};
\item \emph{\bugincorrectcomparison} in four programs, where comparison operators include \textit{==, $<$, and $>$}, etc. 
For example in \textit{QUICKSORT}, the operator \textit{$\geq$} is used instead of \textit{$>$};
\item \emph{\bugline} in four programs. This bug type refers to  buggy programs that miss one or more lines of code. 
For example, the fix for the buggy program \textit{WRAP} is to insert an additional line of code \textit{lines.add(text)}.
\end{inparaenum}
}

This diversity of bug types implies that repair approaches should also consider a diverse set of repair transformations: for example, some bugs could be repaired by replacing operators (e.g., \textit{incorrect comparison operator}); other bugs could be repaired by replacing code (e.g., \textit{ reference  to an incorrect variable}); or by inserting a new line of code (e.g., \textit{missing lines with  a  function  call})).
This implies that, for repairing all of \quixbugs buggy programs, we need one or more approaches capable of applying a wide set of repair transformations.
Thus, if one repair approach can repair most of the buggy programs in \quixbugs, it would mean that this approach is general in essence. 
%\todo{I would remove this last sentence, because a reviewer can say "you have the data, have you found if any approach is general?"}

\paragraph{Program size}
The third column gives the lines of code (LOC) per program ranging from 9 to 67 lines, which can be considered as small. However, we note that 14 are recursive programs and 13 programs contain nested loops. 
It means that, despite a small program size (which can lead to a small search space), the time complexity or space complexity of those programs is sometimes non-trivial.

\paragraph{Characteristics of test suites}
\autoref{tab:quixbug-tests} also summarizes the statistics about JUnit tests: the fourth and fifth columns present the number of passing tests and failing tests. 
As we discussed in \autoref{benchmark}, all programs from the new version of the \quixbugs  have at least one failing JUnit test to expose the bug, which means that the prerequisite of test suite repair is met.
There are 15 programs with no passing tests. All benchmarks of the literature, to our knowledge, contain at least one passing test. 
Passing tests are important for repair approaches to model the expected behavior of the program, 
which means that, without these passing tests, synthesis-based approaches such as Nopol have degenerated synthesis problems when repairing \quixbugs programs.

The sixth column gives the branch coverage information of JUnit tests. 
We observe that the majority of the \quixbugs are completely covered by the test cases (i.e., coverage 100\%).
The least covered program (\emph{MINIMUM\_SPANNING\_TREE}) has a 72\% of coverage.
This high coverage implies that, for most of the branches from the buggy programs, there is at least one test case that executes it.
Thus, any candidate patch applied on those branches will be executed at least once.

\paragraph{Execution time}
The seventh column presents the test execution time for each program. 
There are 37/40 programs whose tests run in less than 2 seconds,  which suggests that program repair will evaluate fast each candidate patch, and eventually repair approaches can completely navigate the search space.
For those 3 programs where the bug triggers a \infiniteloop, the tests timeout after 60 seconds, which explains the 3 large execution time values of programs (\textit{BITCOUNT}, \textit{SQRT}, and \textit{FIND\_FIRST\_IN\_SORTED}).

% failure type
\paragraph{Failure symptoms}
The last column presents the failure symptoms.
We observe 6 different symptoms. 
There are 26 programs with incorrect output failures, 5 programs with \emph{stack overflow failures}, 5 programs with \emph{index out of bounds failures}, 3 programs with \emph{\infiniteloop failures}, 2 programs with \emph{null pointer failures}, and 1 program with \emph{concurrent modification failure}. 
Moreover, we found that two programs, \textit{PASCAL} and \textit{FIND\_FIRST\_IN\_SORTED}, have test cases that expose different failures.
In addition to the diversity of bug types we previously discussed, \quixbugs also has a diversity of test failure symptoms.
This involves that automated program tools must take into account different situations after the bug is executed.
For example, in the case of \emph{\infiniteloop failures}, an approach must avoid handling itself, and in the case of \emph{stack overflow failures} or \emph{index out of bound failures} the repair tool must proceed after the failure and complete the dynamic program analysis such as fault localization.

%programs and preconditions%
\paragraph{Input preconditions}
Preconditions of the input domain for each program are important in our study as we use them as the constraints to automatically generate patches and to discard overfitting patches. 
We computed them for the 40 programs.
All preconditions are given in our online appendix \cite{ourrepo}.
Just to mention one as an example, 
the program 
\textit{GET\_FACTOR} factors an integer value using trial division.
It has a unique function with signature \textit{get\_factor(Integer n): List\textless Integer\textgreater}.
The precondition we found is that the value for integer variable $n$ must be greater than~1. Otherwise, the program is meaningless when the input is a negative integer or zero. And the precondition violated test case generation will influence the patch assessment results in our study.

\begin{figure}[t!]
\includegraphics[width=0.46\textwidth]{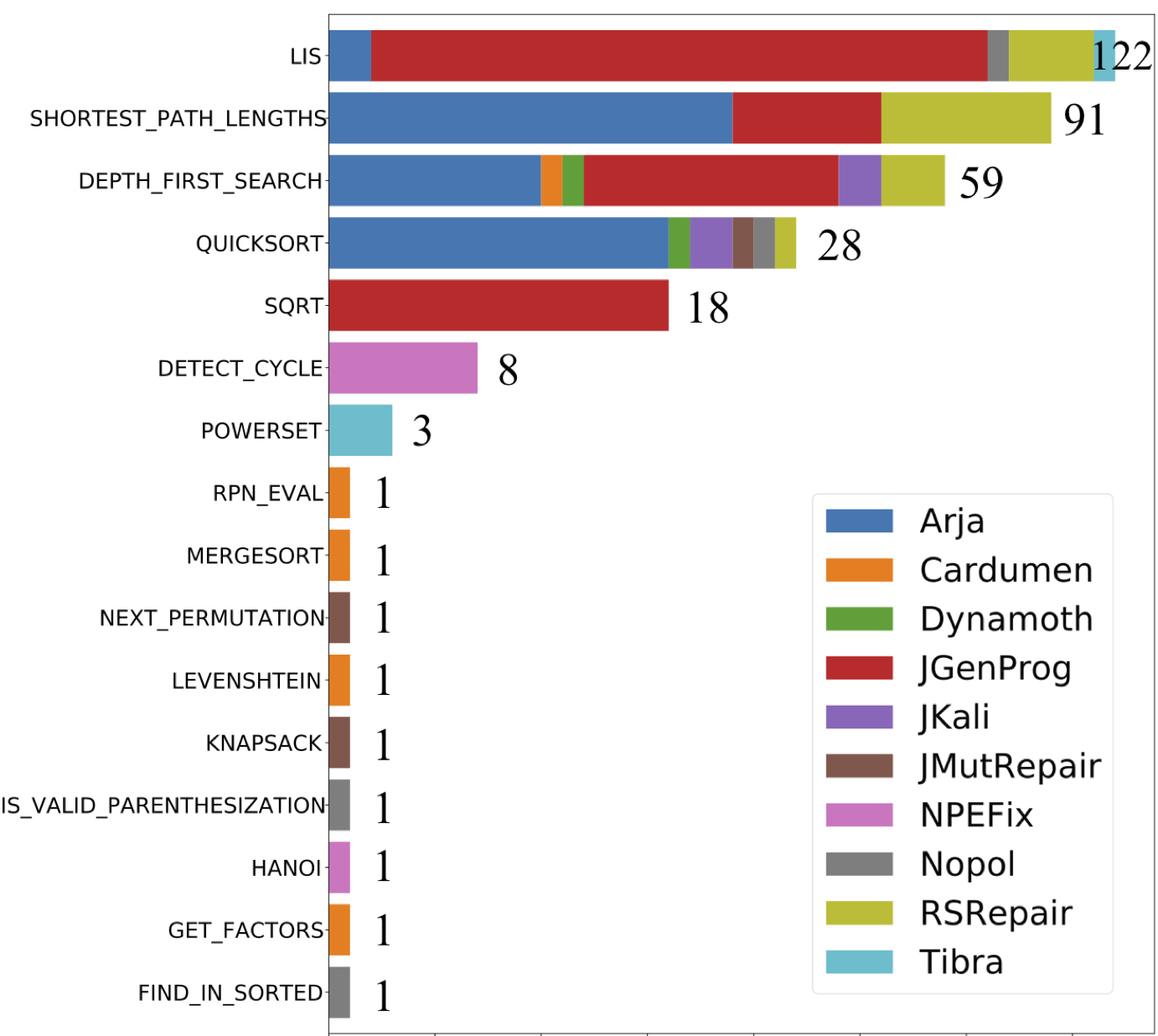}
\caption{The distribution of 338 \quixbugs patches by 10 repair tools.}
\label{repairresult}
\end{figure}

\paragraph{Unique characteristics of \quixbugs}
Comparing the benchmarks of literature \cite{manybugs,introclassjava,defects4jbenchmark,SIR-benchmark,Bears2019,DefextsICSE19,bugjar}, we found  three unique characteristics in \quixbugs{}: 
\begin{inparaenum}[\it 1)]
\item  There is a focus on algorithmic tasks such as \textit{sorting algorithms}, \textit{search algorithms}, \textit{towers of Hanoi puzzle}, whose time complexity or space complexity is non-trivial.
\revision{
The existing benchmarks from the literature (such as Defects4J \cite{defects4jbenchmark} and Bears \cite{Bears2019}) contain buggy programs of real and large open-source libraries, with procedures and modules that implement several functionalities. That is, a buggy program of those benchmarks is not a single textbook algorithm implemented in one single class as each program from QuixBugs is;
}
\item  There are 15/40 programs with only failing tests. To our knowledge, all other benchmarks in the literature always contain at least one passing test;
\item The benchmark contains 3 \infiniteloop failures and 5 stack overflow failures, which is uncommon in bug benchmarks. 
\end{inparaenum} 
Thus, using QuixBugs for program repair will give new insights into the successes and limitations of current repair tools.

%%%summary%%%

\begin{mdframed}
Answer to RQ1: \quixbugs is a valuable dataset for studying program repair.
It has a large diversity of bug types, as well as a diversity of failure symptoms.
It contains buggy programs with unique characteristics compared to existing program repair benchmarks:
\begin{inparaenum}[\it 1)]
\item Complex algorithmic tasks; 
\item Programs with no passing tests; 
\item Programs with a \infiniteloop.
\end{inparaenum}
\end{mdframed}

%%%%%%%%start of patch result%%%%%%%%%%%%% 
%%%%%%%%%%%% RQ2
%%%%%%%%%%%%%%%%%%%%%%%%%%%%%%%%%%%%
\subsection{Results for \RQrepairability}
\label{RQ2repairability}
\begin{table}[t!]
\caption{Number of generated patches per tool.}
\small
\centering
\label{tab:repairresults}
\begin{tabularx}{0.45\textwidth}{@{}Xrr@{}}
\toprule
Repair Tool & \# Patches & \# Repaired Programs\\ 
\midrule
JGenProg   & 164 & 4 \\
Arja       & 113 & 4 \\
RSRepair   & 31 & 3 \\
NPEFix     & 9 & 2 \\
Cardumen   & 5 & 5 \\
Tibra      & 4 & 2 \\
Nopol      & 4 & 4 \\
JKali      & 3 & 2 \\
JMutRepair & 3 & 3 \\
Dynamoth   & 2 & 2 \\
\midrule
Total & 338 patches & 16 patched programs \\
\bottomrule
\end{tabularx}
\end{table}

\revision{
The execution of the ten repair tools produced on the 40 buggy programs of QuixBugs produced \numprint{1470} program repair patches.
Surprisingly, we observe that the repair tools generate duplicated patches.  
We conduct a sanity check and discard syntactically duplicated patches per repair tool. As a result, we discard \numprint{1132} duplicated patches and obtain 338 unique patches.}
\paragraph{Repaired bugs}

We present the results of generated unique patches from our empirical study  in \autoref{repairresult}.
In total, we have obtained 338 patches that repair 16 different buggy programs.
Overall, 40\% of \quixbugs can be repaired with at least one test suite adequate patch.
Note that we have more patches than repaired programs because:
\begin{inparaenum}[\it 1)]
\item Some bugs are repaired by more than one repair tools (e.g., \textit{QUICKSORT}); 
\item Some repair tools generate two or more not duplicated patches for a specific bug (e.g., \textit{Arja} for \textit{QUICKSORT}).
\end{inparaenum}
Note that this empirical study is novel and at scale. To our knowledge, this is the most comprehensive QuixBugs repair empirical study done ever, with the largest number of repair tools and the largest number of patches generated.

\paragraph{Effectiveness of repair tools}
We summarize the effectiveness of repair tools in \autoref{tab:repairresults}.
The first column gives the name of the repair tools. The second and third columns indicate the number of patches generated and the number of programs they repair, respectively.
Cardumen is the approach that repairs the largest number of buggy programs: 5 buggy programs in total can be repaired.
Notably, we observe in \autoref{repairresult} that 4 of them are only repaired by Cardumen.
This shows the extracted code templates in Cardumen are diverse and effective.
Moreover, JGenProg and Arja are two systems that generate the largest number of test suite adequate patches (\numprint{164} and 113 patches).  
This is because JGenProg and Arja leverage multi-objective genetic programming to evolve multiple patches over a series of generations progressively.
There are 4 programs that are repaired by more than three repair tools: \textit{LIS}, \textit{QUICKSORT}, \textit{SHORTEST\_PATH\_LENGTHS} and  \textit{DEPTH\_FIRST\_SEARCH}.
The remaining 12 programs are repaired by only one repair tool.
This implies that specific repair strategies are useful to repair specific bugs. We believe that the global effectiveness of the automatic program repair has to be considered by combining diverse repair approaches together, and not by building a single silver-bullet system.

%bug types%
\paragraph{The repaired bug types} 

\begin{table}[t!]
\small
\caption{Repaired bug types.}
\centering
\label{tab:bugtypesrepaired}
\begin{tabularx}{0.47\textwidth}{@{}lrX@{}}
\toprule
Bug Type (Identified)& \# &  Repaired Programs\\ 
\midrule
{Incorrect comparison operator} & 3/4 &  \textit{knapsack}, \textit{next\_permutation}, \textit{quicksort}\\
{Incorrect arithmetic expression} & 2/2 &  \textit{mergesort}, \textit{sqrt}\\
{Expression swap} & 2/2 &   \textit{shortest\_path\_lengths},  \textit{rpn\_eval} \\
{Missing logic} & 2/3 &  \textit{powerset},  \textit{lis} \\
{Missing ‘+ 1’} & 2/3 &  \textit{find\_in\_sorted}, \textit{levenshtein} \\ 
{Other code replacement} &1/2 &  \textit{is\_valid\_parenthesization} \\ 
{Reference to an incorrect variable} &1/3 & \textit{hanoi}  \\ 
{Wrong constructor call} &1/1 &  \textit{get\_factors} \\ 
{Missing boolean expression} &1/4 & \textit{detect\_cycle}  \\ 
{Missing line with call} &1/4 & \textit{depth\_first\_search}  \\  
\bottomrule
\end{tabularx}
\end{table}

Recall \autoref{tab:quixbug-tests} introduces  17 types of bugs in QuixBugs.
In our empirical study, there are 10/17 types of bugs that are patched by the considered automatic repair approaches. 
We summarize the repair bug types in \autoref{tab:bugtypesrepaired}, where the first column gives the name of the bug type. 
\revision{The second column gives the number of repaired programs belonging to the bug type over the total number of this bug type. We list the repaired programs in the third column.}
The most repaired type of bug is the
\textit{incorrect arithmetic expression}, with 3 programs. 
The 3 repaired  programs were repaired by 6 different repair approaches.
This means that a particular bug type can be repaired using different repair strategies, the reason is that there are different strategies to repair the same bug type.
For example,
Nopol is able to synthesize new statements that use the correct variables instead of the incorrect ones, while JGenProg replaces the buggy statement having the incorrect variable by another one similar that has the correct variable.

In this empirical study, the considered repair tools could not repair 7 bug types:
\begin{inparaenum}[\it 1) ]
\item\bugincorrectlogical;
\item\bugmissingminusone;
\item\bugarray;
\item\bugincorrectmethodcall; 
\item\bugmissingfunctioncall;
\item\bugmissingassignment;
\item\bugmissingarithmetic.
\end{inparaenum}

We now study the reasons for what some types could not be repaired by any approach.
We identify three main reasons:
\begin{inparaenum}[\it 1)]
\item No repair operator implemented;
\item No fixing ingredients; 
\item Limitation of repair implementations.
\end{inparaenum}

{{\emph{No repair operator implemented.}}}
No repair tool repaired the bug type \emph{\bugincorrectlogical} from \textit{BITCOUNT} program.
The ground truth patch updates the a logical operator \textit{``\^~ ''} to \textit{``$\&$''}.
Even if JMutRepair is able to generate patches that change logical and relational operators, it does not implement any mutation of the operator  \textit{``\^~ ''}.

{{\emph{No fixing ingredients.}}}
The advantage of redundancy based repair approaches such as JGenProg, Arja or Cardumen is that they create patches from code already written in the buggy program.
Those approaches could eventually repair a bug of type \bugincorrectlogical if the patch's code (in this case, a binary expression with an operator of \textit{$\&$} ) is present in the buggy program. 
Unfortunately, that is not the case in \textit{BITCOUNT}.
A similar case happens with bug \bugmissingfunctioncall: the ground truth patch for buggy program \textit{SIEVE} replaces a method invocation $any$  by $all$.
However, in that buggy program, there is no piece of code that invokes $all$. 
As a consequence, the redundancy-based repair approaches considered in this empirical study cannot synthesize a fix.

{{\emph{Limitation of repair implementations.}}}
The buggy program \textit{LCS\_LENGTH} that has not one single bug, but two: \textit{incorrect array slice}  and \textit{missing boolean expression}, and the ground truth patch modifies two different locations correspondingly.
Even if, in theory, that buggy program could be repaired by Arja or JGenProg, we observe in practice they could not find a patch because it is a complex multi-location patch.
Multi-location and multi-bug repair are indeed an open research challenge~\cite{8730184}.

%execution time%
\paragraph{Impact of test cases on the capability of  program repair tools} 
For the 16 repaired programs, 5/16 of them have only failing tests and no passing tests.  
To our knowledge, all benchmarks of the literature contain at least one passing test case. Here, our empirical study shows that program repair with only failing tests can be successful. 
There are four programs with no passing tests that are repaired by JGenProg, Arja and RSRepair.
This clearly shows that generate-and-validate repair techniques do not require passing tests for synthesizing a patch. 
\revision{This is because the generate-and-validate repair tools, such as  JGenProg, Arja and RSRepair do not need to infer semantic constraints from passing test cases. They generate the test suite adequate patch through searching the fixing ingredients regardless of passing test cases.
However, not all passing test absent programs can be repaired by generate-and-validate approaches because of the three limitations we have presented above. 
On the contrary, the synthesis-based repair approaches, e.g., Nopol, require passing tests to infer semantic constraints. 
The absence of passing tests creates a degenerated synthesis problem that hampers repair effectiveness.
To this extent, because it contains bugs with no passing test cases, QuixBugs is more appropriate for generate-and-validate repair techniques than for synthesis-based ones.}

% symptom types
\paragraph{On the test failure symptoms of patches programs} 
We have aggregated the failure symptoms of the 16 patched programs: 
10 \textit{incorrect output}, 
4 \textit{stack overflow} errors,
1 \textit{\infiniteloop} error,
and 1 \textit{null pointer exception}.
This confirms the results on Defects4J showing that program repair is effective not only for assertion errors.
No repair approach could repair a bug exposed by a \textit{concurrent modification exception} or an \textit{array index error}.
One possible explanation for this is that test suite based repair tools typically determine suspicious buggy locations based on the root cause of the test failures.  
For those bug types, the test failure symptoms make repair tools difficult to identify the right buggy locations.
For example, the buggy program \textit{FIND\_FIRST\_IN\_SORTED} is a typical incorrect comparison operator bug, which requires the modification of a comparison operator from \textit{``$<=$''} to \textit{``$<$''}, at a location that no repair tools identify as suspicious.  
This suggests the need for alternative fault localization strategies to handle more diverse test failure symptoms, such as pattern-based bug localization \cite{faultloc-patterns}.

\paragraph{On the differences of program repair tools on other benchmarks} 
\revision{
Now, we compare the program repair tools' differences depending on the benchmark, by comparing the repairability over Defects4J and Quixbugs. The considered Defects4J patches are those of Durieux et al. \cite{RepairThemAll2019}, who executed on Defects4J with the same repair tools that we have also considered in this work.
The results are moderately different in repairability rate, failure rate and proportions of duplicated patches.
}

\revision{
First, the repairability rate is the percentage of unique repaired bugs over all bugs from a benchmark. 
The repair tools show a slightly higher repairability rate in Defects4J, i.e., 47.34\% (187/395) \cite{RepairThemAll2019},  than in QuixBugs, i.e., 40\% (16/40).
The higher repairability of Defects4J bugs could be explained, to some extent, in the fact that those bugs are  larger (in LOC) than those from QuixBugs.
This implies that redundancy-based repair approaches (e.g., Cardumen) have  more fixing ingredients available to synthesize a candidate patch, increasing the probability of synthesizing a test suite adequate patch.
}

\revision{
Second, we observe that the repair failure rate -- the percentage of repair attempts that finished due to an error  -- is for QuixBugs (4.31\%) compared to 21.08\% for Defects4J \cite{RepairThemAll2019}. The reason is that Defects4J compared to QuixBugs involves more modules and dependencies during the program execution,  the complexity of Defects4J is higher and hits the limitation of the current automatic patch generation tools. This calls for future research to investigate the implementation of repair tools to mitigate the failure rate.
}

\revision{
Third, in both benchmarks, the 10 considered program repair tools generate a large number of syntactically identical, i.e., duplicated patches, but in different proportions.  Specifically, there exists 51\% (\numprint{19019}/\numprint{37224})  duplicated patches on Defects4J and 77\% (\numprint{1132}/\numprint{1470}) duplicated patches on QuixBugs.  We suspect that the larger number of duplicates on QuixBugs is due to the small size of QuixBugs programs: the amount of fixing ingredients is fewer, those are less diverse than Defects4J, thus they are reused more frequently, and to produce more duplicated patches. 
}

\begin{mdframed}
Answer to RQ2: 16/40 \quixbugs programs are repaired with test suite adequate patches synthesized by ten repair tools. 
Those test suite adequate patches cover 10/17 bug types.
% novelty
A key originality of this empirical study is that it proves that program repair tools work despite the absence of passing tests (five programs without passing tests can be repaired automatically). 
Those results were preliminarily reported in the work \cite{quixbugs}. Here, we report on more patches (338 patches versus 64 patches in \cite{quixbugs}), with unique qualitative insights.
\end{mdframed}
%%%%%End of patch result%%%%%%%%

%%%%%%%%Begin of answer RQ3: patch assessment%%%%

\subsection{\revision{Results for \RQmanualassessment}}
\label{answerrq3}
\begin{figure}[t!]
\centering
\includegraphics[width=0.42\textwidth]{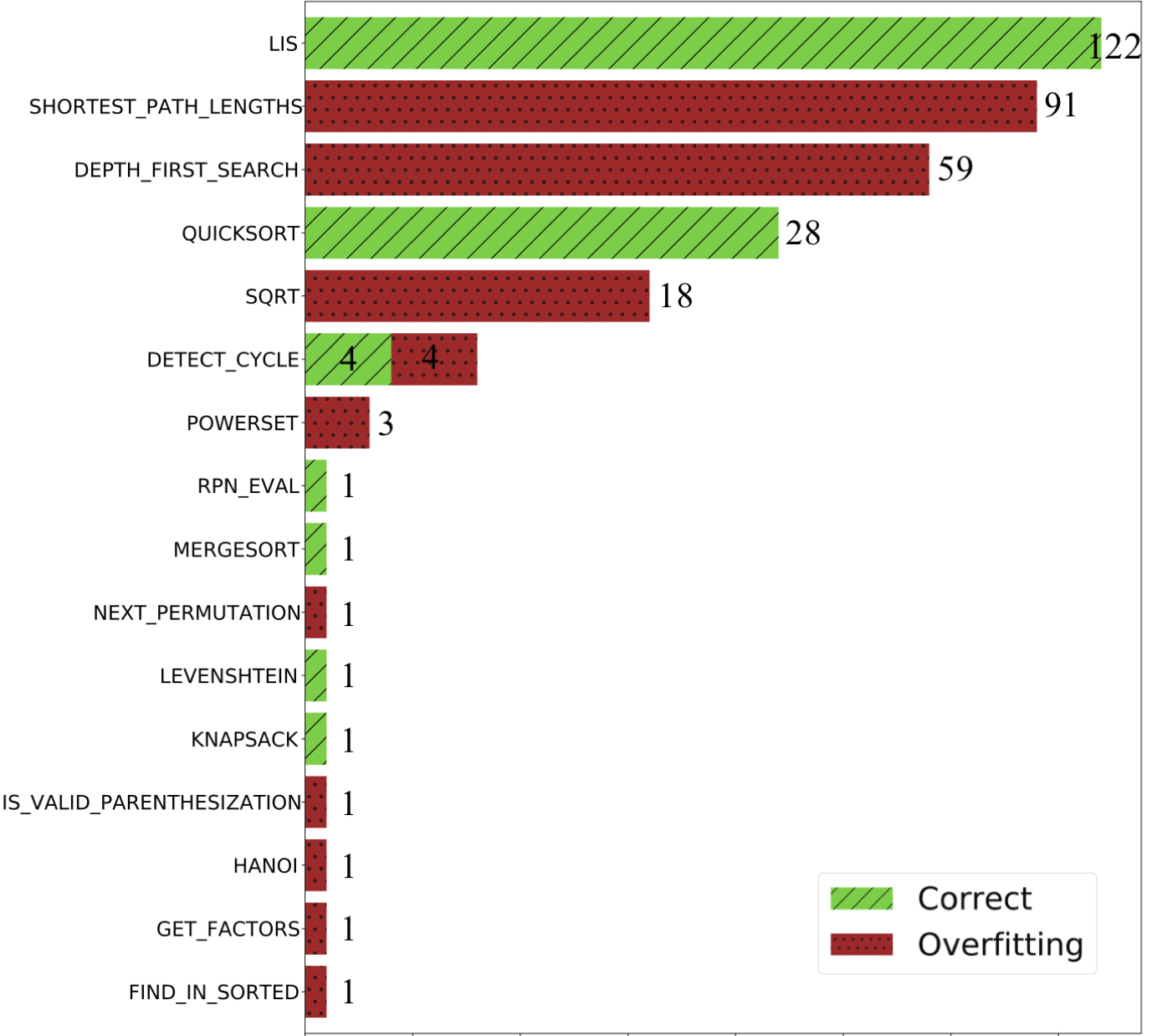}
\caption{Manual assessment of 338 \quixbugs patches spread over 16 QuixBugs bugs.}
\label{fig:manual-analysis-result}
\end{figure}

\revision{
\autoref{fig:manual-analysis-result} shows the manual assessment results for 338 automatic program repair patches synthesized for 16 buggy programs of QuixBugs. 
The green and red legends indicate correct patches and overfitting patches, respectively. 
In total, 158 of 338 are classified as correct, and the remaining 180 patches are classified as overfitting by our manual assessment. 
Overall, per this ground truth based manual assessment, there are 7 out of 40 QuixBugs buggy programs are correctly repaired. 
}

%%% correct repair rate
\revision{
This means that  the repair rate for QuixBugs is 17.5\% (7/40 bugs are correctly repaired), which is respectively 8.1 percentage points and 9.1 percentage points higher than the state-of-the-art evaluations on Defects4J (21/224 bugs are correctly repaired) and IntroClassJava (25/297 bugs are correctly repaired) reported by CapGen \cite{capgen-ICSE18}. 
We explain the higher repair rate for the following two reasons.
First, QuixBugs are small programs, this narrows down the search space of fixing ingredients, and allows for repair tools to precisely locate buggy lines and to find fixing ingredients for patch generation. Second, our empirical study considers more repair tools than previous ones (10 different repair tools) which mechanically increases the number of repaired bugs (\cite{ACS,Yuan2017ARJAAR,Simfix:2018}).
}

%% manual assessment of patch accuracy
\revision{
Now, we talk about the overfitting rate over the generated test suite adequate patches. According to our manual assessment, 53.3\% (180/338) of patches for QuixBugs are overfitting. 
This further confirms that program repair tends to generate more overfitting patches than correct patches \cite{cure-worse-15,Qiissta15}.
Since our empirical study is on a new benchmark, this further strengthens the external validity of this important finding.}

\revision{Notably, we observe that for 15 of 16 buggy programs repaired with test suite adequate patches, either all the generated patches are classified as correct or all overfitting. This suggests that all tools are identically impacted when the specification is weak. Moreover, recall that different repair tools have overlapping repair strategies. For example, Arja and JGenprog are both based on genetic programming search techniques to rearrange the ingredients already existent in the buggy program. 
The outlier is the eight generated patches for the \textit{DETECT\_CYCLE} program, all by NPEFix. 
NPEFix generates eight patches to fix the null pointer exception by adding null checks for variable \textit{hare}. However, we observe that four of those patches, beyond fixing the original bug, introduce a new bug which is not exposed by the original test suite of \textit{DETECT\_CYCLE}.
Thus, we consider those four patches as overfitting, and the other four, which do not suffer that problem, as correct.
}

\begin{figure}[t!]
\centering
\includegraphics[width=0.45\textwidth]{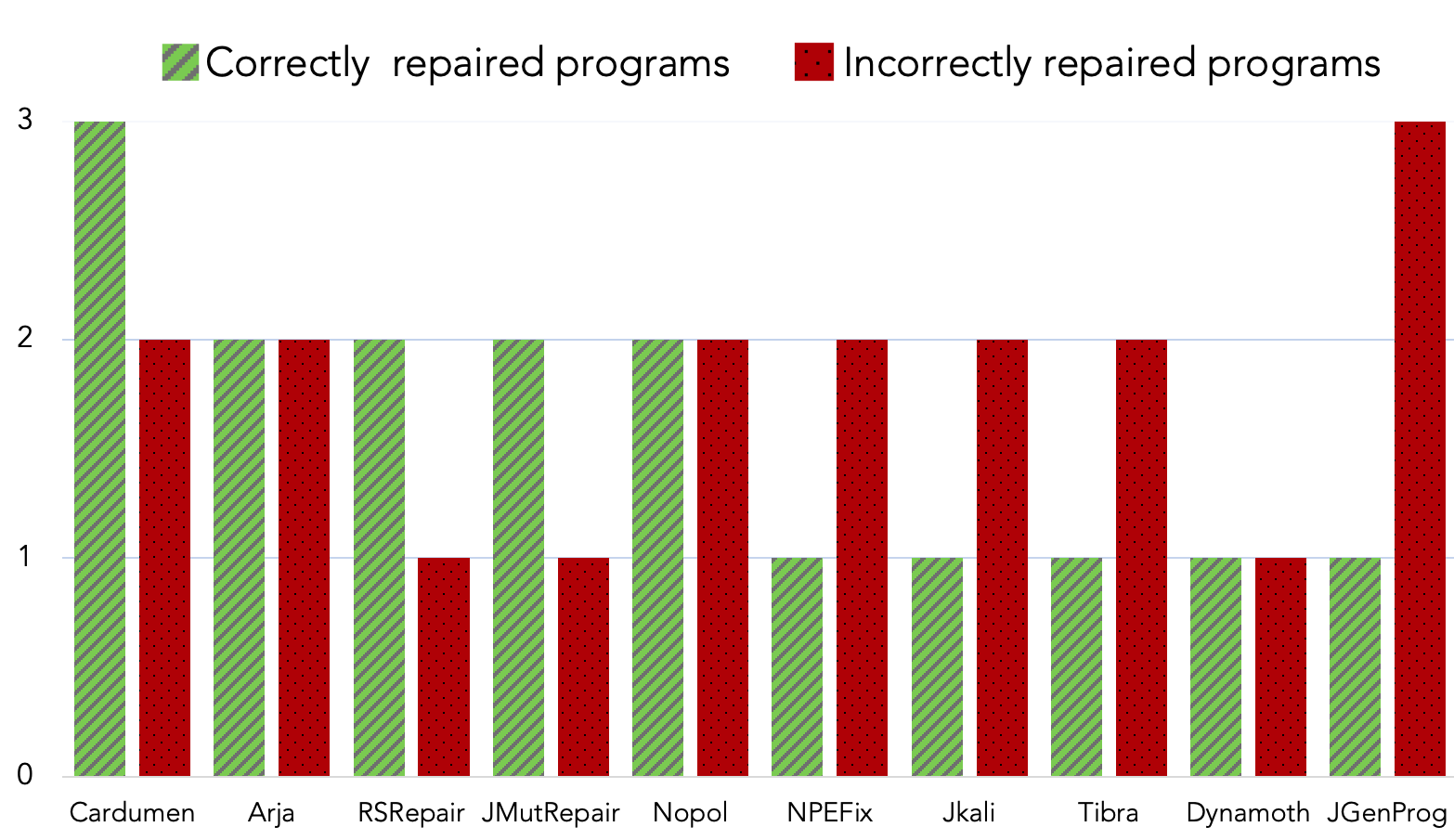}
\caption{The number of buggy programs correctly and incorrectly repaired per repair tool.}
\label{fig:correct-program-per-tool}
\end{figure}

\revision{
\autoref{fig:correct-program-per-tool} presents the number of programs that are correctly repaired per program repair tool. This means the number of programs for which there exists at least one correct patch according to the manual assessment we have done. The green and red legends indicate the number of buggy programs that are respectively correctly and incorrectly repaired per repair tool.
We have the following observations:
\begin{inparaenum}[\it 1)]
\item  All 10 repair tools are able to correctly repair at least one buggy program of QuixBugs;
\item  Cardumen correctly repaired 3 buggy programs, which  outperforms the other 9 repair tools;
\item  Three repair tools contribute to more correctly than incorrectly repaired programs (Cardumen, RSRepair and JMutRepair), three tools perform the same number (Arja, Nopol and Dynamoth), and four tools produce more incorrectly than correctly repaired programs (NPEFix, JKali, Tibra and JGenProg);  
\item Overall, 7 unique programs are correctly repaired by all repair tools, they are complementary to each other.
\end{inparaenum}
}

\revision{
To the best of our knowledge, our study is the first ever to  manually assess 338 patches for QuixBugs. Having this large amount of manually labeled patches is valuable: it paves the way  to  use machine learning techniques to do patch correctness prediction~\cite{staticods}.
}

\begin{mdframed}
\revision{
Answer to RQ3: According to our manual assessment, $7/40$ QuixBugs bugs are correctly repaired, and $158/338$ program repair patches are considered as correct. 
The other $180/338$ generated patches are assessed as overfitting.
Those results are original, because  they are made on a benchmark that is little studied. 
Our findings are aligned to those previously reported on other bug benchmarks: 
\begin{inparaenum}[\it 1)]
\item The number of generated overfitting patches is larger than the number of generated correct patches; 
\item Test suites are too weak to specify program repair patches, even for small programs.
\end{inparaenum}
}
\end{mdframed}

%%%%%%%%Begin of answer RQ4: patch assessment%%%%

%table of rq4
\begin{table*}[t!]
\small
\caption{Number of patches classified as overfitting for QuixBugs programs with at least one patch.}
\centering
\label{tab:rqassessment}
\begin{tabularx}{0.9\textwidth}{@{}>{\scriptsize}Xrrrrrr@{}}
\toprule

\multirow{2}{*}{\small Programs} &\# Generated &\multicolumn{4}{c}{Overfitting patches detected}\\
\cline{3-6}
&patches & $RGT_{Evosuite}$&$RGT_{InputSampling}$&$GT_{Invariants}$&Manual assessment \\

\midrule
LIS &122&0&0&118&0\\
SHORTEST\_PATH\_LENGTHS & 91 & 91 & 91 & 85 &91\\
DEPTH\_FIRST\_SEARCH& 59 &58& 0 & 55 &59 \\
QUICKSORT &28&0&0&3 & 0\\
SQRT & 18  &18& 18 & 18 & 18 \\
DETECT\_CYCLE& 8  &0&0&8 & 4\\
POWERSET &3 &3&3&0 & 3\\
IS\_VALID\_PARENTHESIZATION &1 &1&1&1&1\\
FIND\_IN\_SORTED &1&1&1&1&1\\
HANOI &1&1&1&0&1\\
GET\_FACTORS& 1 &1&0&0&1\\

NEXT\_PERMUTATION &1&0&0&0&1\\
RPN\_EVAL &1&0&0&0&0\\
KNAPSACK& 1&0&0&0&0\\
LEVENSHTEIN &1&0&0&0&0\\
MERGESORT &1&0&0&0&0\\
\midrule
{\small Sum} & 338   & 174 & 115 & 289&180 \\
\bottomrule
\end{tabularx}
\end{table*}

\subsection{Results for \RQcorrectnessassessment}

\revision{In this section, we analyze the results of automated patch correctness assessments.
We compare them against the results obtained from the manual assessment.  
This comparison allows us to calculate the accuracy of the considered automated assessment techniques.
Finally, we discuss their true/false positive and negative cases.
}

\revision{\subsubsection{RQ4a: Patches classified as overfitting by automated patch assessments}}
\autoref{tab:rqassessment} shows the overfitting patch assessment results produced by the three considered techniques over 338 generated patches. 
The first column gives the names of buggy programs patched by at least one repair tool. 
The second column shows the total number of generated patches over all tools.
The third to fifth columns give the number of patches classified as overfitting  by the three automated assessments. We present the manual assessment results in the last column.
%overfitting patches found by the corresponding techniques. 
\revision{For example, the first row shows there are 122 patches generated for bug \textit{LIS}, $RGT_{Evosuite}$ and $RGT_{InputSampling}$ identify all of them as correct patches (i.e., 0 overfitting patch), which is fully aligned with manual assessment results. However, the $GT_{Invariants}$ classifies 118 of them as overfitting, which contradicts the manual assessment. } 

\revision{
The automated patch assessment techniques classify patches as overfitting with different magnitude:
$RGT_{Evosuite}$ identifies 174/338 (51.5\%) patches from 8 buggy programs as overfitting, 
$RGT_{InputSampling}$ identifies 115/338 (34\%) patches from 6 buggy programs as overfitting, and 
$GT_{Invariants}$ identifies 289/338 (85.5\%) patches  from 8 buggy programs as overfitting. 
The differences between those numbers can be explained by the kind of information they collect, and are influenced by outliers.
For example, $GT_{Invariants}$ classifies more overfitting patches than the two other techniques, this is mostly because it classifies 118 patches for program \textit{LIS} as overfitting while the other two techniques do not.
On the contrary, $RGT_{InputSampling}$, which classifies as overfitting less patches than the other two techniques, 
does not classify as overfitting any patch from program \textit{DEPTH\_FIRST\-\_SEARCH}, while the other two techniques do it for at least 55 patches.
}

%Together
\revision{
Now, we compare the results from automated assessment with those from manual assessment, which classifies as overfitting 180 patches.
That comparison allows us to detect the misclassification of the automated assessments.
}

\begin{figure}[t!]
\centering
\includegraphics[width=7cm]{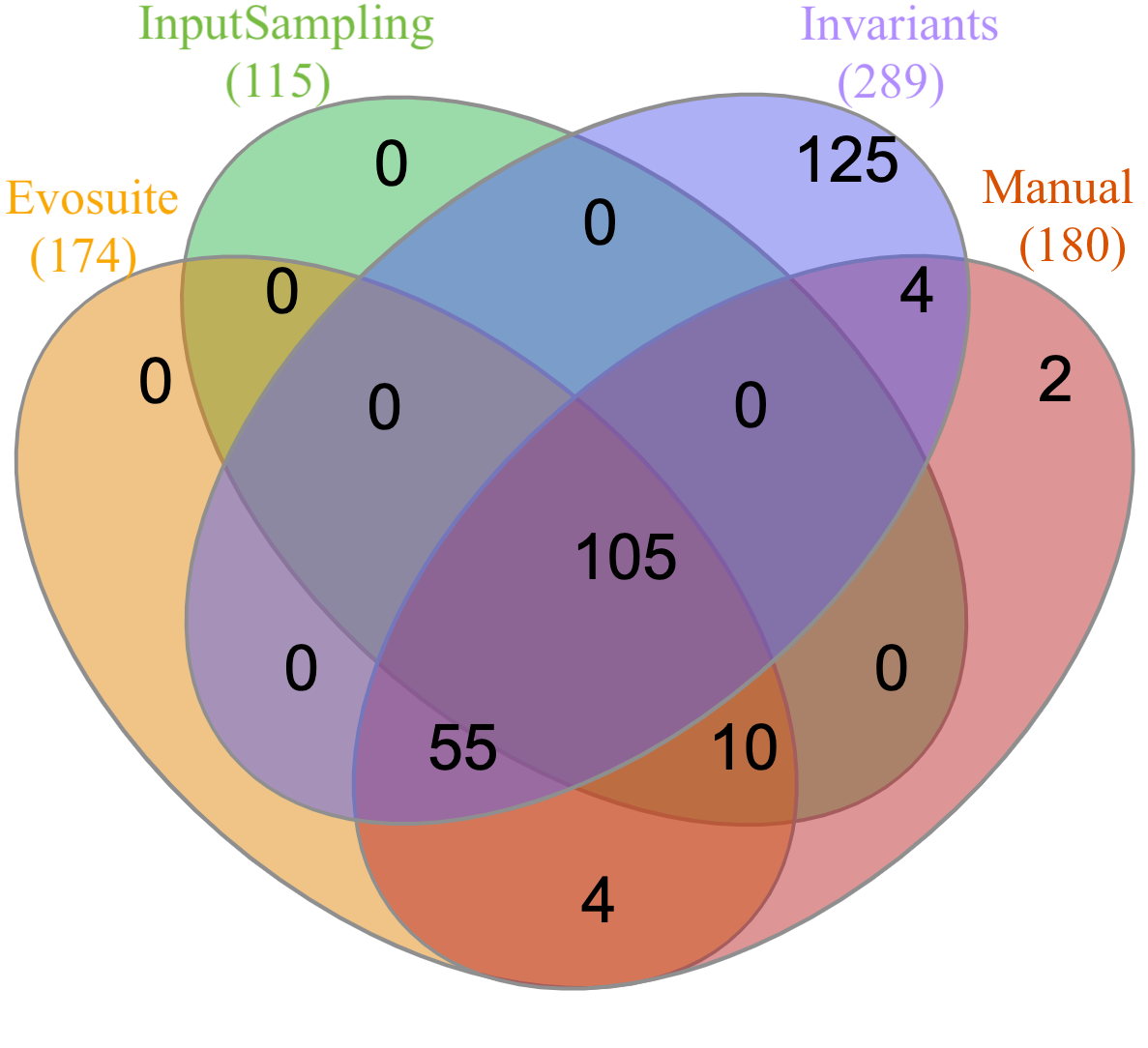}
\caption{Venn diagram showing the overlap between the three considered automated assessment techniques and manual assessment.}
\label{fig:rqassessmentresultVenn}
\end{figure}

\revision{
The Venn diagram in \autoref{fig:rqassessmentresultVenn} shows the overlap between the different assessments.
First, we observe that the manual assessment and the three techniques agree on the majority (105) of overfitting patches (the intersection of all circles). 
This shows that all considered automated assessment techniques can correctly identify, at least,  58.3\% (105/180) of overfitting patches. 
%overlap
Second, the assessment result from $RGT_{Evosuite}$ is the closest to manual assessment with 174 of 180 cases. Overall, $RGT_{Evosuite}$ is able to correctly classify the 174/180 overfitting patches.
%overlap
Third, all overfitting patches classified by $RGT_{InputSampling}$ can be found by $RGT_{Evosuite}$. 
The overlap between them is due to the similarity of the techniques, both based on the generation of test inputs.
However, $RGT_{Evosuite}$ is able to correctly classify 59 additional patches.
Fourth, $GT_{Invariants}$ classifies as overfitting the largest number of patches (289). However, that is due to the misclassification of 125 correct patches.
}

% consensus
\revision{
The three automated techniques achieve patch correctness assessment results consensus on the majority of buggy programs (9/16 programs). 
For three buggy programs, they agree on the overfitting diagnostic for all patches, for which our
manual assessment also classified them as overfitting.
Moreover, the three techniques achieve consensus on the absence of overfitting patches for 5 buggy programs (the last 5 rows of  \autoref{tab:rqassessment}).  
However, according to our manual assessment, there is one patch for program \textit{NEXT\_PERMUTATION} which is actually overfitting. % NEXT\_PERMUTATION 
}

\begin{table}[t!]
\caption{Accuracy of the three automated patch assessments. }
\small
\centering
\label{tab:falsepositives}
\begin{tabular}{@{}lrrrrr@{}}
\toprule
Automated Assessment & \# TP & \#FP  & \# TN& \# FN & Accuracy \\ 
\midrule
$RGT_{Evosuite}$ &174  & 0  & 158 & 6 &  98.2\%\\
$RGT_{InputSampling}$ & 115 & 0  & 158 & 65  & 80.8\% \\
$GT_{Invariants}$ & 164 & 125  & 33 & 16 & 58.3\%\\

\bottomrule
\end{tabular}
\end{table}

%%%%%%%%%%%%%bnginning of rq4 %%%%%%%%%%%%%%

\revision{\subsubsection{RQ4b: Accuracy of automated patch assessments}}
%introduce manual assessmeng result.
%table of true negatives and false positives
\revision{
The manual assessment enables us to compute the accuracy of the three automated correctness assessments.
}

{{\textbf{Accuracy.}}}
\autoref{tab:falsepositives} gives the results of this analysis.
%describle table
The first column gives the name of the automated assessment technique, and the second to fourth columns indicate the number of true positives, true negatives, false positives, and false negatives. The individual accuracy is given in the last column according to \autoref{equation:accuracy}.
For instance, $RGT_{Evosuite}$ yields 174 true positives, 0 false positive,  158 true negatives, and 6 false negatives. 
This means there are 174 and 158 patches that are correctly classified as overfitting and correct, respectively.  On the contrary, $RGT_{Evosuite}$ fails to identify 6 overfitting patches.

The $RGT_{Evosuite}$, $RGT_{InputSampling}$, and $GT_{Invariants}$  respectively achieve an accuracy rate of 98.2\%, 80.8\%, and 58.3\%. 
%accuracy
$RGT_{Evosuite}$ achieves the best accuracy among the three assessment techniques. 
This is inline with previous results having shown that Evosuite performs better than pure random test generation \cite{Shamshiriase15,drr,automatic-test-competition}.
Also, $RGT_{Evosuite}$ has better accuracy than $GT_{Invariants}$, this is because  $RGT_{Evosuite}$ produces fewer false positives than $GT_{Invariants}$ (0 versus 125).

%true negatives and false negatives.
\textbf{True and false negatives.}
$RGT_{Evosuite}$ and $RGT_{InputSampling}$ report the same number of true negatives cases (158), which are correct patches not classified as overfitting. 
Now, we see that all three techniques have false negatives, which are overfitting patches classified as correct.
In the case of test generation techniques such as $RGT_{Evosuite}$ and $RGT_{InputSampling}$,
the false negative cases appear when the generated tests do not contain inputs that expose the incorrect behavior of an overfitting patch. 
$RGT_{InputSampling}$ produces the most false negative cases (65). This shows the weaker effectiveness of the domain-specific generators developer to sample test inputs.
Finally, no technique is able to identify all overfitting patches and have zero false negatives.

\begin{listing}[t!]
\tiny
\caption{An overfitting patch by JMutRepair that is only found by manual assessment.}
\label{list-nextper}
\begin{lstlisting}
      for (int j=perm.size()-1; j!=i; j--) {
    -   if ((perm.get(j)) < (perm.get(i))) {
    +   if ((perm.get(j)) >= (perm.get(i))) {
        //ground truth patch:
        //if (perm.get(j) > perm.get(i)) {
\end{lstlisting}
          
\end{listing}

% case study of false negative
We discuss a case of an overfitting patch that is not identified as such by any automated techniques. % $RGT_{Evosuite}$,  $RGT_{InputSampling}$ and $GT_{Invariants}$.
\autoref{list-nextper} gives the patch of  \textit{NEXT\_PERMUTATION} generated by JMutRepair.
The bug is present in an if condition which  compares the position of two elements in a list \textit{perm}.
The generated patch uses the \textit{``$>=$''} operator to fix the bug. 
This patch is not identical to the ground truth version, which uses the operator \textit{``$>$''} instead. 
The manual assessment reveals that this patch is overfitting: if the list \textit{perm} contains the same values, the behaviors of the JMutRepair patch and the ground truth patch differ and the programs produce different outputs.
The aforementioned specific input was neither generated by $RGT_{Evosuite}$ nor $RGT_{InputSampling}$.
Consequently, the JMutRepair patch is not classified as overfitting.  
This case study illustrates that automated correctness assessment cannot fully replace manual assessment.

\begin{listing*}[ht!]
\tiny
\caption{A case study of a false positive of $GT_{Invariants}$.}
\noindent\begin{minipage}[b]{0.48\textwidth}
\begin{lstlisting}
public static ArrayList<Integer> quicksort(ArrayList<Integer> arr) {
if (arr.isEmpty()) {
    return new ArrayList<Integer>();
}
Integer pivot = arr.get(0);

ArrayList<Integer> lesser = new ArrayList<Integer>();
ArrayList<Integer> greater = new ArrayList<Integer>();

for (Integer x : arr.subList(1, arr.size())) {
if (x < pivot) {
    lesser.add(x);
<@\color{red}{-    \} else if (x > pivot) \{ //buggy code }@>
<@\color{greencomments}{+   \} else if (x >= pivot) \{ //ground truth patch}@>
    greater.add(x);
}

    \end{lstlisting}
    \captionof{sublisting}{The ground-truth patch of QUICKSORT}
    \label{invar-ground-truth}            
    \end{minipage}
    \hfill
    \hfill
\begin{minipage}[b]{0.48\textwidth}
    \begin{lstlisting}
Program: QUICKSORT
(java.util.ArrayList):::EXIT	
===============================================
// always hold
arr[] == orig(arr[])	
return != null	
return[] elements != null	
return[].getClass().getName() elements == java.lang.Integer.class	
//the following invariants in gray hold for the ground truth patch and listing2c patch, but not listing2d 
<@\colorbox{lightgray}{(size(arr[])-1==-1)==>(arr[] == [])	}@>
<@\colorbox{lightgray}{(size(arr[])-1==-1)==>(arr[].getClass().getName()==[])	}@>
<@\colorbox{lightgray}{(size(arr[])-1==-1)==>(return[] == [])}@>	
<@\colorbox{lightgray}{(size(arr[])-1==-1)==>(return[].getClass().getName()==[])}@>
\end{lstlisting}
\captionof{sublisting}{Invariants captured from the ground truth program execution}
\label{invariant-listing}            
\end{minipage}
\begin{minipage}[b]{0.48\textwidth}
    \begin{lstlisting}[firstnumber=13]
public static ArrayList<Integer> quicksort(ArrayList<Integer> arr){
    if (arr.isEmpty()) {
<@\color{red}{-  ~~~    return new ArrayList<Integer>();}@>
<@\color{greencomments}{+  ~~   if (arr.isEmpty()) \{}@>
<@\color{greencomments}{+   ~~~   return new ArrayList<Integer>();}@>
<@\color{greencomments}{+  ~~   \}}@>
   }
Integer pivot = arr.get(0);
ArrayList<Integer> lesser = new ArrayList<Integer>();
ArrayList<Integer> greater = new ArrayList<Integer>();
for (Integer x : arr.subList(1, arr.size())) {
    if (x < pivot) {
        lesser.add(x);
<@\color{red}{-   ~~ \} else if (x > pivot) \{ @>
<@\color{greencomments}{+  ~~  \} else  \{ @>
      greater.add(x);
   }
    \end{lstlisting}
 \captionof{sublisting}{A semantically equivalent patch  classified as correct by $GT_{Invariants}$}
 \label{invar-ground-correct}              
\end{minipage}
 \hfill
\begin{minipage}[b]{0.48\textwidth}
    \begin{lstlisting}
public static ArrayList<Integer> quicksort(ArrayList<Integer> arr){
   if (arr.isEmpty()) {
     return new ArrayList<Integer>();
   }
   Integer pivot = arr.get(0);
<@\color{greencomments}{+ ~  if (arr.isEmpty()) \{}@>
<@\color{greencomments}{+ ~~~ return new ArrayList<Integer>();}@>
<@\color{greencomments}{+ ~ \}}@>
ArrayList<Integer> lesser = new ArrayList<Integer>();
ArrayList<Integer> greater = new ArrayList<Integer>();
for (Integer x : arr.subList(1, arr.size())) {
  if (x < pivot) {
    lesser.add(x);
<@\color{red}{-   \} else if (x > pivot) \{ }@>
<@\color{greencomments}{+   \} else \{  }@>
    greater.add(x);
   }
\end{lstlisting}
\captionof{sublisting}{A semantically equivalent patch classified as overfitting by $GT_{Invariants}$}
\label{invar-ground-overfitting}            
\end{minipage}
\label{invariant-case-study}  
\end{listing*}

%true positives and false positives.
{{\textbf{True and false positives.}}}
\autoref{tab:falsepositives} shows that $RGT_{Evosuite}$ has the largest number of true positive cases among the three assessment techniques. 
$RGT_{Evosuite}$ classifies 59 and 10 more overfitting patches than $RGT_{InputSampling}$ and $GT_{Invariants}$.
All overfitting patches classified by $RGT_{Evosuite}$  and  $RGT_{InputSampling}$  are true positive cases (i.e., they do not suffer any false positive case). Those two approaches are 100\% precise.   
On the contrary, $GT_{Invariants}$ suffers from a large number of false positive cases (125). 
Thus, the precision of this technique is 56.7\% (164/289).
We identify two main reasons behind those false positives.
First,  the invariants may capture a specific value of the test case, instead of capturing the full range. For instance, an invariant may capture \textit{a.value == 0} instead of  \textit{a.value $>=$ 0} because only 0 is used in the test case. In this case, the invariant itself is overfitting and results in a false positive.
Second, invariants detection is sensitive to procedure exit and entry points, where the preconditions and postconditions are obtained from.  
When a patched program adds new exit points (e.g., new return statements), all invariants that hold for the ground truth program are expected to hold at new exit points, otherwise, a patch is assessed as a violation, and this is the major reason for those false positives. 

% False positives with Invariants
% \emph{Case study of false positive.}
\autoref{invariant-case-study} is an example of false positive for $GT_{Invariants}$.
\autoref{invar-ground-truth} is the ground truth patch for the buggy program \textit{QUICKSORT}. 
The invariants captured from the ground truth program execution are given in \autoref{invariant-listing}. 
% explanation of the patch
That ground truth patch modifies an operator, from \textit{``$>$''} to \textit{``$>=$''}.

Our empirical study found two patches for this QuixBugs subject: one presented in \autoref{invar-ground-correct} classified as correct by $GT_{Invariants}$, and a second one  presented in \autoref{invar-ground-overfitting} classified as overfitting by the same assessment.
Both patches modify the conditions of \textit{else} block,  using ``\textit{else}`` to replace ``\textit{else if (x $>$ pivot)}``.
These changes are semantically equivalent to those proposed by the ground truth patch, so the patches are correct.

We note that these two generated patches add redundant statements of \textit{if (arr.isEmpty()}, which do not influence the correctness of the patches.
However, those  statements impact the correctness evaluation done by $GT_{Invariants}$ because they introduce new exit points (new return statements).
In \autoref{invariant-listing}, 
we can see that an invariant states a property for variable \textit{arr} at the method exit point. 
The invariants in gray hold for the patch in  \autoref{invar-ground-correct}, because the exit point \textit{return new  ArrayList\textless Integer\textgreater();} meets all captured invariants, all four properties captured for variable \textit{arr} (e.g., \textit{arr[] == []}).
However, in the patch at \autoref{invar-ground-overfitting}, a new program exit point is added one line after the first exit point, and one invariant is not satisfied.
When the \textit{arr} is not empty, the program enters into the new exit point, and invariant \textit{arr[] == []} is violated.

% implications
To sump, capturing behavioral differences with invariants violations is hard. In particular, invariants that hold at a certain point in a program typically no longer hold in the patched program when new exit points that are added. 
This is an important caveat for assessing patches generated by genetic programming (e.g., JGenProg and Arja), which tend to generate many new statements with exit points.  
This suggests interesting future research directions to improve $GT_{Invariants}$.

\begin{mdframed}
Answer to RQ4:  
The accuracy of $RGT_{Evosuite}$, $RGT_{InputSampling}$ and $GT_{Invariants}$ are 98.2\%, 80.8\% and 58.3\%, respectively, showing that $RGT_{Evosuite}$ is the best patch assessment technique on \quixbugs. 
To our knowledge, this is the first study ever that uses automated patch assessment on \quixbugs, showing both its feasibility and effectiveness. 
We note that $GT_{Invariants}$, proposed by \cite{ibf20-invariant}, suffers from many false positives, calling for more research on this topic, little explored by the program repair research community.
\end{mdframed}

%%%%%%%%%%%%%%%%%%%%%%%%%%%%%%%%%%%%%%%%%%%%%%%%%%
%% NEW SECTION
%%%%%%%%%%%%%%%%%%%%%%%%%%%%%%%%%%%%%%%%%%%%%%%%%%
\section{Threats to Validity}
\label{threat}

The major threat in our study lies in the manual correctness assessment, which may result in misclassification due to a lack of expertise or mistakes.
This threat holds for all program repair results based on manual assessment. The best mitigation to this threat is to make patches and analyses  publicly available. Then other researchers are able to further assess them, this is what we have done in our open-science repository \cite{ourrepo}.

We note that manual assessment done over \quixbugs{} patches is less error-prone than for Defects4J and more complex benchmarks because:
\begin{inparaenum}[\it 1)]
\item \quixbugs contains well-known algorithms (e.g., \textit{QUICKSORT}), thus the patch analyst does not need to be an expert in the buggy program's application domain;
\item As presented in \autoref{tab:repairresults}, \quixbugs programs are short, thus it is easier to read, debug and understand the generated patches.
\end{inparaenum}

The second threat is about construct validity. 
The automatic repair tools that we used in this empirical study could have bugs that prevent them to discover all possible patches.
For this reason, the results we have reported are likely an under-estimation of the repairability of \quixbugs using automatic program repair. 

\revision{
\section{Discussion}
\label{discussisons} 
\paragraph{The benefits of using QuixBugs} Studying QuixBugs provides two major benefits for the research community. 
First, it enables the community to make new findings about program repair that have never been reported in other benchmarks, that we will discuss next. 
Second, it strengthens the external validity of the findings previously found on other benchmarks.}

\revision{
The study presented in this paper enables us to identify the following new findings:
\begin{inparaenum}[\it 1)]
\item Generate-and-validate approaches are capable of generating patches for buggy programs with only failing test cases, while synthesis-based approaches cannot. The other benchmarks do not contain buggy programs with only failing tests, thus this finding has never been reported before;
\item 7 of 40 (17.5\%) buggy programs of QuixBugs are correctly repaired, which is 8.1\% and 9.1\%  higher than for Defects4J and IntroClassJava benchmarks, respectively (see \autoref{answerrq3});
Our explanation is that, as the QuixBugs programs are smaller than for other benchmarks, the corresponding search spaces are also smaller, thus, the repair tools are able to navigate a bigger portion of the search space, increasing the probability of finding the  patch;
\item The automated patch assessment technique  $RGT_{Evosuite}$ has the best effectiveness in our study, it is able to identify 98.2\% overfitting patches. 
\end{inparaenum}
}

\revision{
Our novel empirical study on QuixBugs confirms the following findings found on other benchmarks, showing their generalizability:
1) Our study on QuixBugs confirms the existence of overfitting patches by a large amount; 
2) Our study on QuixBugs confirms that both generate-and-validate approaches and synthesis-based approaches work, but on different bugs.
}

\revision{
\paragraph{The potential future improvement of program repair tools}
Although the repair rate of QuixBugs is higher than the existing benchmarks of Defects4J and IntroClassJava,  still 33/40 of buggy programs are not able to be repaired with correct patches. 
For instance, some bugs that could be repaired with a simple one-liner fix (e.g., GCD) are not able to be automatically repaired by any repair tool.
Our empirical study suggests potential future improvements of program repair tools. }

\revision{
First, we observe the generate-and-validate approaches (e.g., Arja and JGenProg) based on genetic programming, commonly generating complex repair code.
This results in redundant code being repeatedly used in multiple locations of the buggy programs, which are not necessary for repairing the bugs. The redundant and complex codes lead to the difficulty in patch understanding for researchers and developers. 
This has been discussed by Yuan and Banzhaf \cite{Yuan2017ARJAAR}, we advocate the future improvement in the simplicity of repair code into the search process of genetic programming algorithms.
}

\revision{
Second, we observe that program repair tools tend to generate high-granularity patches, which results in the abstract syntax tree (AST) edit operations often appearing in the root nodes of the buggy statements.
These coarse-grained repair operations are not effective for repairing finer-grained bugs, such as those that can be repaired by swapping two variables or replacing incorrect reference variables (e.g., \textit{BUCKETSORT} and \textit{GCD}).
Thus, we consider lower-granularity patch generation will benefit future program repair research.
}

\revision{
Third, 
program repair tools could benefit from  more diverse fixing ingredients, including 
\begin{inparaenum}[\it 1)]
\item New operators: for example, in \autoref{RQ2repairability} we explained the absence of operators for repairing \textit{BITCOUNT}; 
\item Repair patterns: some considered repair tools such as JGenProg and Cardumen only consider ingredients taken from the buggy program, which not enough to repair a bug;
\item New code synthesis mechanism:  for example, a repair tool could use a new mechanism to create patches with  \emph{visible} and \emph{invisible} method invocations, The visible and invisible method invocations are respectively referring to the method that exists in the buggy programs and widely used utility packages (e.g., Commons Math project).  In the considered 10 repair tools, we do not observe any  patches with an invisible method invocation.
\end{inparaenum}
}

\section{Related Work}

\label{relatedwork}
\subsection{Datasets of Bugs}
The benchmarks used in automatic program repair research include  Introclass \cite{manybugs}, ManyBugs \cite{manybugs}, SIR \cite{SIR-benchmark}, Codeflaws \cite{Tancodeflaws},
Defects4J \cite{defects4jbenchmark}, IntroClassJava \cite{introclassjava}, Bugs.jar \cite{bugjar},  Bears \cite{Bears2019}, and Defexts \cite{DefextsICSE19}.

%papers use introclass benchmark
Smith et al. \cite{cure-worse-15} evaluate overfitting patches generated by GenProg and TrpAutoRepair on IntroClass. 
Le et al. \cite{LeICSE18} systematically characterize the nature of overfitting in semantics-based automatic program repair on the IntroClass and Codeflaws benchmarks.
Ke et al. \cite{kease15} evaluate SearchRepair on IntroClass. 
% \emph{ManyBugs:} 
Qi et al. \cite{Qiissta15}, Mechtaev et al. \cite{Angelixicse16}, Long and Rinard \cite{SPR-manybugs}  evaluate repair approaches on ManyBugs. Stratis and Rajan \cite{Stratis-ase16}  evaluate their approach on SIR.  Nguyen et al. \cite{semfix} propose SemFix and evaluate it on SIR.
Papadakis et al. \cite{mutant} collect and analyze mutant quality indicators based on Codeflaws. Chekam et al.~\cite{mutanticse18} propose a new technique for mutant selection and evaluate their work on Codeflaws.
%paper use defects4j benchmark
Martinez et al. \cite{Defects4JExperiment} and  Yu et al. \cite{zhongxingpaper} report on their experiments using Defects4J.
Xiong et al. \cite{ACS} propose the ACS repair tool and evaluate it on four projects of the Defects4J benchmark. Wen et al. \cite{capgen-ICSE18} propose CapGen, a context-aware patch generation technique and evaluate this technique on the Defects4J.  Hua et al. \cite{sketchfix} propose and evaluate SketchFix on Defects4J.
% all program repair papers using IntroClassJava as benchmark
Wen et al. \cite{capgen-ICSE18} and
Le et al. \cite{LeS3fse17} propose and evaluate their repair technique on IntroClassJava.

% multi benchmark
The  recent  work  by Durieux  et  al. \cite{RepairThemAll2019} conducted a large scale empirical study on five benchmarks, Defects4J, Bears, IntroClassJava, Bugs.jar and QuixBugs. However, they do not provide any assessment for the generated patches.

To the best of our knowledge, our study is the first ever to assess 338 patches for \quixbugs.

\subsection{Patch Correctness Assessment}
Synthesizing new inputs for patch correctness assessment has been studied in a few papers. Xin and Risse \cite{issta17}  propose DiffTGen to identify overfitting patches with tests generated by Evosuite \cite{evosuite}.
Those tests are meant to detect behavioral differences between a generated patch and a human-written patch. 

Xiong et al. \cite{patch-correctness-icse18}  propose  PATCH-SIM  and TEST-SIM to heuristically determine patch correctness by comparing the execution similarity of the original and newly generated tests before and after the patch. 

Yang et al. \cite{Opad} propose Opad and Gao et al. \cite{ISSTA19GaoCrash} propose Fix2Fit, two approaches based on implicit oracles for detecting overfitting patches that introduce crashes or memory-safety problems.
These two approaches for automatic patch correctness assessment cannot be applied to QuixBugs which contain Java programs with no low-level memory problems.

Tan et al. \cite{anti-pattern} aim to identify the overfitting patches with predefined templates to capture typical overfitting behaviors. While their approach is static, our approach is dynamic. 
In our study, the test inputs are executed and the invariants are captured from program runtime behavior in order to detect overfitting patches.

Yang and Yang \cite{ibf20-invariant} study invariants generation to infer behaviors of generated patches. Their study shows that the majority of plausible patches (92/96) expose different runtime behaviors.  
Our study also considers invariants based patch assessment, but at a much larger scale: first, our dataset is  three times larger (338 versus 96 patches) and second, we also measure the accuracy and false positives which have not been done in \cite{ibf20-invariant}.

\section{Conclusion}
\label{conclusion}

We have presented a novel program repair empirical study, studying the \quixbugs{} benchmark \cite{quixbugs-orig} and 10 repair tools. We have compared three automated patch assessment techniques over 338 generated patches. Lastly, we have comprehensively studied the accuracy and false positives of the three considered assessment techniques.
%external validity 
Our empirical study yields major findings for program repair research: 
\begin{inparaenum}[\it 1)]
\item It is possible to repair programs with no passing tests at all (only failing test cases); 
\item Patch assessment with $RGT_{Evosuite}$ has the highest accuracy over the three considered techniques in discarding overfitting patches.
 \end{inparaenum}
Finally, our empirical study results in 338 patches with correctness labels, which is a valuable asset for future study on  overfitting in automatic program repair.

\section*{Acknowledgments}
We thank the anonymous reviewers for their insightful comments on early draft of the paper.
This work was supported by the Wallenberg AI, Autonomous Systems and Software Program (WASP) funded by the Knut and Alice Wallenberg Foundation.

\bibliographystyle{elsarticle-num}
\bibliography{references}

\begin{thebibliography}{10}
\expandafter\ifx\csname url\endcsname\relax
  \def\url#1{\texttt{#1}}\fi
\expandafter\ifx\csname urlprefix\endcsname\relax\def\urlprefix{URL }\fi
\expandafter\ifx\csname href\endcsname\relax
  \def\href#1#2{#2} \def\path#1{#1}\fi

\bibitem{genprog}
C.~{Le Goues}, T.~V. Nguyen, S.~Forrest, W.~Weimer, {GenProg: A generic method
  for automatic software repair}, in: IEEE Transactions on Software
  Engineering, Vol.~38, 2012, pp. 54--72.
\newblock \href {http://dx.doi.org/10.1109/TSE.2011.104}
  {\path{doi:10.1109/TSE.2011.104}}.

\bibitem{Astor2019Journal}
M.~Martinez, M.~Monperrus, Astor: Exploring the design space of
  generate-and-validate program repair beyond genprog, Journal of Systems and
  Software 151 (2019) 65 -- 80.
\newblock \href {http://dx.doi.org/https://doi.org/10.1016/j.jss.2019.01.069}
  {\path{doi:https://doi.org/10.1016/j.jss.2019.01.069}}.

\bibitem{capgen-ICSE18}
M.~Wen, J.~Chen, R.~Wu, D.~Hao, S.~C. Cheung, {Context-aware patch generation
  for better automated program repair}, in: Proceedings - International
  Conference on Software Engineering, Vol. 2018-January of ICSE '18, 2018, pp.
  1--11.
\newblock \href {http://dx.doi.org/10.1145/3180155.3180233}
  {\path{doi:10.1145/3180155.3180233}}.

\bibitem{autofix}
Y.~Pei, C.~A. Furia, M.~Nordio, Y.~Wei, B.~Meyer, A.~Zeller, {Automated fixing
  of programs with contracts}, in: IEEE Transactions on Software Engineering,
  Vol.~40 of ISSTA '10, 2014, pp. 427--449.
\newblock \href {http://arxiv.org/abs/1403.1117} {\path{arXiv:1403.1117}},
  \href {http://dx.doi.org/10.1109/TSE.2014.2312918}
  {\path{doi:10.1109/TSE.2014.2312918}}.

\bibitem{semfix}
H.~D.~T. Nguyen, D.~Qi, A.~Roychoudhury, S.~Chandra, {SemFix: Program repair
  via semantic analysis}, in: Proceedings - International Conference on
  Software Engineering, 2013, pp. 772--781.
\newblock \href {http://dx.doi.org/10.1109/ICSE.2013.6606623}
  {\path{doi:10.1109/ICSE.2013.6606623}}.

\bibitem{Nopol}
J.~Xuan, M.~Martinez, F.~DeMarco, M.~Clement, S.~L. Marcote, T.~Durieux, D.~{Le
  Berre}, M.~Monperrus, {Nopol: Automatic Repair of Conditional Statement Bugs
  in Java Programs}, IEEE Transactions on Software Engineering 43~(1) (2017)
  34--55.
\newblock \href {http://arxiv.org/abs/1811.04211} {\path{arXiv:1811.04211}},
  \href {http://dx.doi.org/10.1109/TSE.2016.2560811}
  {\path{doi:10.1109/TSE.2016.2560811}}.

\bibitem{martinbibli}
M.~Monperrus, {Automatic software repair: A bibliography}, ACM Computing
  Surveys 51~(1).
\newblock \href {http://arxiv.org/abs/1807.00515} {\path{arXiv:1807.00515}},
  \href {http://dx.doi.org/10.1145/3105906} {\path{doi:10.1145/3105906}}.

\bibitem{TSE-repair-survey}
L.~Gazzola, D.~Micucci, L.~Mariani, {Automatic Software Repair: A Survey}, IEEE
  Transactions on Software Engineering 45~(1) (2019) 34--67.
\newblock \href {http://dx.doi.org/10.1109/TSE.2017.2755013}
  {\path{doi:10.1109/TSE.2017.2755013}}.

\bibitem{manybugs}
C.~{Le Goues}, N.~Holtschulte, E.~K. Smith, Y.~Brun, P.~Devanbu, S.~Forrest,
  W.~Weimer, {The ManyBugs and IntroClass Benchmarks for Automated Repair of C
  Programs}, in: IEEE Transactions on Software Engineering, Vol.~41, 2015, pp.
  1236--1256.
\newblock \href {http://dx.doi.org/10.1109/TSE.2015.2454513}
  {\path{doi:10.1109/TSE.2015.2454513}}.

\bibitem{defects4jbenchmark}
R.~Just, D.~Jalali, M.~D. Ernst, {Defects4J: A database of existing faults to
  enable controlled testing studies for Java programs}, in: 2014 International
  Symposium on Software Testing and Analysis, ISSTA 2014 - Proceedings, 2014,
  pp. 437--440.

\bibitem{sketchfix}
J.~Hua, M.~Zhang, K.~Wang, S.~Khurshid, {Towards practical program repair with
  on-demand candidate generation}, in: ICSE, 2018, pp. 12--23.
\newblock \href {http://dx.doi.org/10.1145/3180155.3180245}
  {\path{doi:10.1145/3180155.3180245}}.

\bibitem{patch-correctness-icse18}
Y.~Xiong, X.~Liu, M.~Zeng, L.~Zhang, G.~Huang, {Identifying patch correctness
  in test-based program repair}, in: Proceedings - International Conference on
  Software Engineering, 2018, pp. 789--799.
\newblock \href {http://arxiv.org/abs/1706.09120} {\path{arXiv:1706.09120}},
  \href {http://dx.doi.org/10.1145/3180155.3180182}
  {\path{doi:10.1145/3180155.3180182}}.

\bibitem{quixbugs-orig}
D.~Lin, A.~Chen, J.~Koppel, A.~Solar-Lezama, {QuixBugs: A Multi-Lingual Program
  Repair Benchmark Set Based on the Quixey Challenge}, in: SPLASH Companion
  2017 - Proceedings Companion of the 2017 ACM SIGPLAN International Conference
  on Systems, Programming, Languages, and Applications: Software for Humanity,
  2017, pp. 55--56.
\newblock \href {http://dx.doi.org/10.1145/3135932.3135941}
  {\path{doi:10.1145/3135932.3135941}}.

\bibitem{Yuan2017ARJAAR}
Y.~Yuan, W.~Banzhaf, {ARJA: Automated Repair of Java Programs via
  Multi-Objective Genetic Programming}, in: IEEE Transactions on Software
  Engineering, 2018, pp. 1--1.
\newblock \href {http://arxiv.org/abs/1712.07804} {\path{arXiv:1712.07804}},
  \href {http://dx.doi.org/10.1109/TSE.2018.2874648}
  {\path{doi:10.1109/TSE.2018.2874648}}.

\bibitem{cardumen}
M.~Martinez, M.~Monperrus, {Ultra-large repair search space with automatically
  mined templates: The cardumen mode of astor}, in: Lecture Notes in Computer
  Science (including subseries Lecture Notes in Artificial Intelligence and
  Lecture Notes in Bioinformatics), Vol. 11036 LNCS, 2018, pp. 65--86.
\newblock \href {http://arxiv.org/abs/1712.03854} {\path{arXiv:1712.03854}},
  \href {http://dx.doi.org/10.1007/978-3-319-99241-9_3}
  {\path{doi:10.1007/978-3-319-99241-9_3}}.

\bibitem{dynamoth}
T.~Durieux, M.~Monperrus, {DynaMoth: Dynamic code synthesis for automatic
  program repair}, in: Proceedings - 11th International Workshop on Automation
  of Software Test, AST 2016, 2016, pp. 85--91.
\newblock \href {http://dx.doi.org/10.1145/2896921.2896931}
  {\path{doi:10.1145/2896921.2896931}}.

\bibitem{NPEFixAR}
B.~Cornu, T.~Durieux, L.~Seinturier, M.~Monperrus,
  \href{http://arxiv.org/abs/1512.07423}{{NPEFix: Automatic Runtime Repair of
  Null Pointer Exceptions in Java}}, CoRR 2015\href
  {http://arxiv.org/abs/1512.07423} {\path{arXiv:1512.07423}}.
\newline\urlprefix\url{http://arxiv.org/abs/1512.07423}

\bibitem{Qiissta15}
Z.~Qi, F.~Long, S.~Achour, M.~Rinard, {An analysis of patch plausibility and
  correctness for generate-and-validate patch generation systems}, in: 2015
  International Symposium on Software Testing and Analysis, ISSTA 2015 -
  Proceedings, 2015, pp. 24--36.
\newblock \href {http://dx.doi.org/10.1145/2771783.2771791}
  {\path{doi:10.1145/2771783.2771791}}.

\bibitem{overfit-problem}
Q.~Xin, {Towards addressing the patch overfitting problem}, in: Proceedings -
  2017 IEEE/ACM 39th International Conference on Software Engineering
  Companion, ICSE-C 2017, 2017, pp. 489--490.
\newblock \href {http://dx.doi.org/10.1109/ICSE-C.2017.42}
  {\path{doi:10.1109/ICSE-C.2017.42}}.

\bibitem{cure-worse-15}
E.~K. Smith, E.~T. Barr, C.~{Le Goues}, Y.~Brun, {Is the cure worse than the
  disease? Overfitting in automated program repair}, in: 2015 10th Joint
  Meeting of the European Software Engineering Conference and the ACM SIGSOFT
  Symposium on the Foundations of Software Engineering, ESEC/FSE 2015 -
  Proceedings, 2015, pp. 532--543.
\newblock \href {http://dx.doi.org/10.1145/2786805.2786825}
  {\path{doi:10.1145/2786805.2786825}}.

\bibitem{LeICSE18}
X.~B.~D. Le, F.~Thung, D.~Lo, C.~L. Goues, {Overfitting in semantics-based
  automated program repair}, Empirical Software Engineering 23~(5) (2018)
  3007--3033.
\newblock \href {http://dx.doi.org/10.1007/s10664-017-9577-2}
  {\path{doi:10.1007/s10664-017-9577-2}}.

\bibitem{Defects4JExperiment}
M.~Martinez, T.~Durieux, R.~Sommerard, J.~Xuan, M.~Monperrus, {Automatic repair
  of real bugs in java: a large-scale experiment on the defects4j dataset}, in:
  Empirical Software Engineering, Vol.~22, 2017, pp. 1936--1964.
\newblock \href {http://arxiv.org/abs/1811.02429} {\path{arXiv:1811.02429}},
  \href {http://dx.doi.org/10.1007/s10664-016-9470-4}
  {\path{doi:10.1007/s10664-016-9470-4}}.

\bibitem{ourrepo}
H.~Ye, M.~Martinez, T.~Durieux, M.~Monperrus, {Github repository of our study
  is available online}, \url{https://github.com/KTH/quixbugs-experiment}
  (2020).

\bibitem{quixbugs}
H.~Ye, M.~Martinez, T.~Durieux, M.~Monperrus, {A Comprehensive Study of
  Automatic Program Repair on the QuixBugs Benchmark}, in: IBF 2019 - 2019 IEEE
  1st International Workshop on Intelligent Bug Fixing, 2019, pp. 1--10.
\newblock \href {http://arxiv.org/abs/1805.03454} {\path{arXiv:1805.03454}},
  \href {http://dx.doi.org/10.1109/IBF.2019.8665475}
  {\path{doi:10.1109/IBF.2019.8665475}}.

\bibitem{ibf20-invariant}
B.~Yang, J.~Yang, {Exploring the Differences between Plausible and Correct
  Patches at Fine-Grained Level}, in: 2020 IEEE 2nd International Workshop on
  Intelligent Bug Fixing (IBF), 2020, pp. 1--8.
\newblock \href {http://dx.doi.org/10.1109/ibf50092.2020.9034821}
  {\path{doi:10.1109/ibf50092.2020.9034821}}.

\bibitem{quixey-challenge}
R.~Lawler, {How do you hire great engineers? Give them a challenge},
  \url{https://gigaom.com/2012/01/19/quixey-challenge/} (2012).

\bibitem{martinICSE14}
M.~Monperrus, {A critical review of "automatic patch generation learned from
  human-written patches": Essay on the problem statement and the evaluation of
  automatic software repair}, in: Proceedings - International Conference on
  Software Engineering, no.~1 in ICSE'14, 2014, pp. 234--242.
\newblock \href {http://arxiv.org/abs/1408.2103} {\path{arXiv:1408.2103}},
  \href {http://dx.doi.org/10.1145/2568225.2568324}
  {\path{doi:10.1145/2568225.2568324}}.

\bibitem{survey-bugs}
T.~Zhang, H.~Jiang, X.~Luo, A.~T. Chan, {A Literature Review of Research in Bug
  Resolution: Tasks, Challenges and Future Directions}, The Computer Journal
  59~(5) (2016) 741--773.
\newblock \href {http://dx.doi.org/10.1093/comjnl/bxv114}
  {\path{doi:10.1093/comjnl/bxv114}}.

\bibitem{relation-bug-cause}
Z.~Ni, B.~Li, X.~Sun, T.~Chen, B.~Tang, X.~Shi, Analyzing bug fix for automatic
  bug cause classification, Journal of Systems and Software 163 (2020) 110538.
\newblock \href {http://dx.doi.org/https://doi.org/10.1016/j.jss.2020.110538}
  {\path{doi:https://doi.org/10.1016/j.jss.2020.110538}}.

\bibitem{RepairThemAll2019}
T.~Durieux, F.~Madeiral, M.~Martinez, R.~Abreu, {Empirical review of Java
  program repair tools: A large-scale experiment on 2,141 bugs and 23,551
  repair attempts}, in: ESEC/FSE 2019 - Proceedings of the 2019 27th ACM Joint
  Meeting European Software Engineering Conference and Symposium on the
  Foundations of Software Engineering, 2019, pp. 302--313.
\newblock \href {http://arxiv.org/abs/1905.11973} {\path{arXiv:1905.11973}},
  \href {http://dx.doi.org/10.1145/3338906.3338911}
  {\path{doi:10.1145/3338906.3338911}}.

\bibitem{issta17-difftgen}
Q.~Xin, S.~P. Reiss, {Identifying test-suite-overfitted patches through test
  case generation}, in: ISSTA 2017 - Proceedings of the 26th ACM SIGSOFT
  International Symposium on Software Testing and Analysis, 2017, pp. 226--236.
\newblock \href {http://dx.doi.org/10.1145/3092703.3092718}
  {\path{doi:10.1145/3092703.3092718}}.

\bibitem{ACS}
Y.~Xiong, J.~Wang, R.~Yan, J.~Zhang, S.~Han, G.~Huang, L.~Zhang, {Precise
  condition synthesis for program repair}, in: Proceedings - 2017 IEEE/ACM 39th
  International Conference on Software Engineering, ICSE 2017, 2017, pp.
  416--426.
\newblock \href {http://arxiv.org/abs/1608.07754} {\path{arXiv:1608.07754}},
  \href {http://dx.doi.org/10.1109/ICSE.2017.45}
  {\path{doi:10.1109/ICSE.2017.45}}.

\bibitem{drr}
H.~Ye, M.~Martinez, M.~Monperrus, {Automated Patch Assessment for Program
  Repair at Scale} (2019).
\newblock \href {http://arxiv.org/abs/1909.13694} {\path{arXiv:1909.13694}}.

\bibitem{zhongxingpaper}
Z.~Yu, M.~Martinez, B.~Danglot, T.~Durieux, M.~Monperrus, {Alleviating patch
  overfitting with automatic test generation: a study of feasibility and
  effectiveness for the Nopol repair system}, Empirical Software Engineering
  24~(1) (2019) 33--67.
\newblock \href {http://arxiv.org/abs/1810.10614} {\path{arXiv:1810.10614}},
  \href {http://dx.doi.org/10.1007/s10664-018-9619-4}
  {\path{doi:10.1007/s10664-018-9619-4}}.

\bibitem{randomtest}
A.~Arcuri, M.~Z. Iqbal, L.~Briand, {Random testing: Theoretical results and
  practical implications}, IEEE Transactions on Software Engineering 38~(2)
  (2012) 258--277.
\newblock \href {http://dx.doi.org/10.1109/TSE.2011.121}
  {\path{doi:10.1109/TSE.2011.121}}.

\bibitem{issta17}
Q.~Xin, S.~P. Reiss, {Identifying test-suite-overfitted patches through test
  case generation}, in: ISSTA 2017 - Proceedings of the 26th ACM SIGSOFT
  International Symposium on Software Testing and Analysis, 2017, pp. 226--236.
\newblock \href {http://dx.doi.org/10.1145/3092703.3092718}
  {\path{doi:10.1145/3092703.3092718}}.

\bibitem{Opad}
J.~Yang, A.~Zhikhartsev, Y.~Liu, L.~Tan, {Better test cases for better
  automated program repair}, in: Proceedings of the ACM SIGSOFT Symposium on
  the Foundations of Software Engineering, Vol. Part F130154, 2017, pp.
  831--841.
\newblock \href {http://dx.doi.org/10.1145/3106237.3106274}
  {\path{doi:10.1145/3106237.3106274}}.

\bibitem{evosuite}
G.~Fraser, A.~Arcuri, {EvoSuite: Automatic test suite generation for
  object-oriented software}, in: SIGSOFT/FSE 2011 - Proceedings of the 19th ACM
  SIGSOFT Symposium on Foundations of Software Engineering, 2011, pp. 416--419.
\newblock \href {http://dx.doi.org/10.1145/2025113.2025179}
  {\path{doi:10.1145/2025113.2025179}}.

\bibitem{Shamshiriase15}
S.~Shamshiri, R.~Just, J.~M. Rojas, G.~Fraser, P.~McMinn, A.~Arcuri, {Do
  automatically generated unit tests find real faults? An empirical study of
  effectiveness and challenges}, in: Proceedings - 2015 30th IEEE/ACM
  International Conference on Automated Software Engineering, ASE 2015, 2016,
  pp. 201--211.
\newblock \href {http://dx.doi.org/10.1109/ASE.2015.86}
  {\path{doi:10.1109/ASE.2015.86}}.

\bibitem{randomness-guide}
A.~Arcuri, L.~Briand, {A Hitchhiker's guide to statistical tests for assessing
  randomized algorithms in software engineering}, in: Software Testing
  Verification and Reliability, Vol.~24 of ICSE '11, 2014, pp. 219--250.
\newblock \href {http://dx.doi.org/10.1002/stvr.1486}
  {\path{doi:10.1002/stvr.1486}}.

\bibitem{mechtaev2018semantic}
S.~Mechtaev, M.-D. Nguyen, Y.~Noller, L.~Grunske, A.~Roychoudhury, {Semantic
  program repair using a reference implementation}, in: Proceedings of ICSE,
  2018, pp. 129--139.
\newblock \href {http://dx.doi.org/10.1145/3180155.3180247}
  {\path{doi:10.1145/3180155.3180247}}.

\bibitem{daikon}
M.~D. Ernst, J.~H. Perkins, P.~J. Guo, S.~McCamant, C.~Pacheco, M.~S. Tschantz,
  C.~Xiao, {The Daikon system for dynamic detection of likely invariants},
  Science of Computer Programming 69~(1-3) (2007) 35--45.
\newblock \href {http://dx.doi.org/10.1016/j.scico.2007.01.015}
  {\path{doi:10.1016/j.scico.2007.01.015}}.

\bibitem{introclassjava}
T.~Durieux, M.~Monperrus, Introclassjava: A benchmark of 297 small and buggy
  java programs, Tech. rep., Universite Lille 1 (2016).

\bibitem{SIR-benchmark}
H.~Do, S.~Elbaum, G.~Rothermel, {Supporting controlled experimentation with
  testing techniques: An infrastructure and its potential impact}, Empirical
  Software Engineering 10~(4) (2005) 405--435.
\newblock \href {http://dx.doi.org/10.1007/s10664-005-3861-2}
  {\path{doi:10.1007/s10664-005-3861-2}}.

\bibitem{Bears2019}
F.~Madeiral, S.~Urli, M.~Maia, M.~Monperrus, {BEARS: An Extensible Java Bug
  Benchmark for Automatic Program Repair Studies}, in: SANER 2019 - Proceedings
  of the 2019 IEEE 26th International Conference on Software Analysis,
  Evolution, and Reengineering, IEEE, Hangzhou, China, 2019, pp. 468--478.
\newblock \href {http://arxiv.org/abs/1901.06024} {\path{arXiv:1901.06024}},
  \href {http://dx.doi.org/10.1109/SANER.2019.8667991}
  {\path{doi:10.1109/SANER.2019.8667991}}.

\bibitem{DefextsICSE19}
S.~Benton, A.~Ghanbari, L.~Zhang, {Defexts: A curated dataset of reproducible
  real-world bugs for modern JVM languages}, in: Proceedings - 2019 IEEE/ACM
  41st International Conference on Software Engineering: Companion,
  ICSE-Companion 2019, ICSE '19, 2019, pp. 47--50.
\newblock \href {http://dx.doi.org/10.1109/ICSE-Companion.2019.00035}
  {\path{doi:10.1109/ICSE-Companion.2019.00035}}.

\bibitem{bugjar}
R.~K. Saha, Y.~Lyu, W.~Lam, H.~Yoshida, M.~R. Prasad, {Bugs.jar: A large-scale,
  diverse dataset of real-world Java bugs}, in: Proceedings - International
  Conference on Software Engineering, MSR '18, Association for Computing
  Machinery, New York, NY, USA, 2018, pp. 10--13.
\newblock \href {http://dx.doi.org/10.1145/3196398.3196473}
  {\path{doi:10.1145/3196398.3196473}}.

\bibitem{8730184}
B.~{Khaireddine}, M.~{Martinez}, A.~{Mili}, Program repair at arbitrary fault
  depth, in: 2019 12th IEEE Conference on Software Testing, Validation and
  Verification (ICST), 2019, pp. 465--472.

\bibitem{faultloc-patterns}
X.~Sun, W.~Zhou, B.~Li, Z.~Ni, J.~Lu,
  \href{https://doi.org/10.1109/ACCESS.2019.2894976}{Bug localization for
  version issues with defect patterns}, {IEEE} Access 7 (2019) 18811--18820.
\newblock \href {http://dx.doi.org/10.1109/ACCESS.2019.2894976}
  {\path{doi:10.1109/ACCESS.2019.2894976}}.
\newline\urlprefix\url{https://doi.org/10.1109/ACCESS.2019.2894976}

\bibitem{Simfix:2018}
J.~Jiang, Y.~Xiong, H.~Zhang, Q.~Gao, X.~Chen, {Shaping program repair space
  with existing patches and similar code}, in: ISSTA 2018 - Proceedings of the
  27th ACM SIGSOFT International Symposium on Software Testing and Analysis,
  ISSTA, 2018, pp. 298--309.
\newblock \href {http://dx.doi.org/10.1145/3213846.3213871}
  {\path{doi:10.1145/3213846.3213871}}.

\bibitem{staticods}
H.~Ye, J.~Gu, M.~Martinez, T.~Durieux, M.~Monperrus,
  \href{http://arxiv.org/pdf/1910.12057}{Automated classification of
  overfitting patches with statically extracted code features}, Tech. Rep.
  1910.12057, arXiv (2019).
\newline\urlprefix\url{http://arxiv.org/pdf/1910.12057}

\bibitem{automatic-test-competition}
F.~Kifetew, X.~Devroey, U.~Rueda, {Java Unit Testing Tool Competition - Seventh
  Round}, in: Proceedings - 2019 IEEE/ACM 12th International Workshop on
  Search-Based Software Testing, SBST 2019, SBST '19, 2019, pp. 15--20.
\newblock \href {http://dx.doi.org/10.1109/SBST.2019.00014}
  {\path{doi:10.1109/SBST.2019.00014}}.

\bibitem{Tancodeflaws}
S.~H. Tan, J.~Yi, Yulis, S.~Mechtaev, A.~Roychoudhury, {Codeflaws: A
  programming competition benchmark for evaluating automated program repair
  tools}, in: Proceedings - 2017 IEEE/ACM 39th International Conference on
  Software Engineering Companion, ICSE-C 2017, 2017, pp. 180--182.
\newblock \href {http://dx.doi.org/10.1109/ICSE-C.2017.76}
  {\path{doi:10.1109/ICSE-C.2017.76}}.

\bibitem{kease15}
Y.~Ke, K.~T. Stolee, C.~L. Goues, Y.~Brun, {Repairing programs with semantic
  code search}, in: Proceedings - 2015 30th IEEE/ACM International Conference
  on Automated Software Engineering, ASE 2015, 2016, pp. 295--306.
\newblock \href {http://dx.doi.org/10.1109/ASE.2015.60}
  {\path{doi:10.1109/ASE.2015.60}}.

\bibitem{Angelixicse16}
S.~Mechtaev, J.~Yi, A.~Roychoudhury, {Angelix: Scalable multiline program patch
  synthesis via symbolic analysis}, in: Proceedings - International Conference
  on Software Engineering, Vol. 14-22-May-2016, 2016, pp. 691--701.
\newblock \href {http://dx.doi.org/10.1145/2884781.2884807}
  {\path{doi:10.1145/2884781.2884807}}.

\bibitem{SPR-manybugs}
F.~Long, M.~Rinard, {Staged program repair with condition synthesis}, in: 2015
  10th Joint Meeting of the European Software Engineering Conference and the
  ACM SIGSOFT Symposium on the Foundations of Software Engineering, ESEC/FSE
  2015 - Proceedings, 2015, pp. 166--178.
\newblock \href {http://dx.doi.org/10.1145/2786805.2786811}
  {\path{doi:10.1145/2786805.2786811}}.

\bibitem{Stratis-ase16}
P.~Stratis, A.~Rajan, {Test case permutation to improve execution time}, in:
  ASE 2016 - Proceedings of the 31st IEEE/ACM International Conference on
  Automated Software Engineering, 2016, pp. 45--50.
\newblock \href {http://dx.doi.org/10.1145/2970276.2970331}
  {\path{doi:10.1145/2970276.2970331}}.

\bibitem{mutant}
P.~Mike, T.~T. Chekam, Y.~{Le Traon}, {Mutant quality indicators}, Proceedings
  - 2018 IEEE 11th International Conference on Software Testing, Verification
  and Validation Workshops, ICSTW 2018 (2018) 32--39\href
  {http://dx.doi.org/10.1109/ICSTW.2018.00025}
  {\path{doi:10.1109/ICSTW.2018.00025}}.

\bibitem{mutanticse18}
T.~T. Chekam, M.~Papadakis, T.~Bissyand{\'{e}}, Y.~L. Traon, {Poster:
  Predicting the fault revelation utility of mutants}, in: Proceedings -
  International Conference on Software Engineering, 2018, pp. 408--409.
\newblock \href {http://dx.doi.org/10.1145/3183440.3195031}
  {\path{doi:10.1145/3183440.3195031}}.

\bibitem{LeS3fse17}
X.~B.~D. Le, D.~H. Chu, D.~Lo, C.~{Le Goues}, W.~Visser, {S3: Syntax- and
  semantic-guided repair synthesis via programming by examples}, in:
  Proceedings of the ACM SIGSOFT Symposium on the Foundations of Software
  Engineering, Vol. Part F130154, 2017, pp. 593--604.
\newblock \href {http://dx.doi.org/10.1145/3106237.3106309}
  {\path{doi:10.1145/3106237.3106309}}.

\bibitem{ISSTA19GaoCrash}
X.~Gao, S.~Mechtaev, A.~Roychoudhury, {Crash-avoiding program repair}, in:
  ISSTA 2019 - Proceedings of the 28th ACM SIGSOFT International Symposium on
  Software Testing and Analysis, ISSTA 2019, 2019, pp. 8--18.
\newblock \href {http://dx.doi.org/10.1145/3293882.3330558}
  {\path{doi:10.1145/3293882.3330558}}.

\bibitem{anti-pattern}
S.~H. Tan, H.~Yoshida, M.~R. Prasad, A.~Roychoudhury, Anti-patterns in
  search-based program repair, in: Proceedings of the 2016 24th ACM SIGSOFT
  International Symposium on Foundations of Software Engineering, FSE 2016,
  Association for Computing Machinery, New York, NY, USA, 2016, p. 727–738.
\newblock \href {http://dx.doi.org/10.1145/2950290.2950295}
  {\path{doi:10.1145/2950290.2950295}}.

\end{thebibliography}

\end{document}